\documentclass[apj]{emulateapj}
\slugcomment{{\sc Accepted to ApJS:} April 19, 2010}

\usepackage{graphicx}
\usepackage{epstopdf}
\usepackage{color}
\usepackage{multirow}

\usepackage{calrsfs}
\usepackage{units}
\usepackage{lscape}

\usepackage{natbib}
\newcommand{\solmy}{\>{\rm M_{\odot}yr^{-1}}}

\newcommand{\as}{^{\prime\prime}}
\newcommand{\am}{^{\prime}}
\newcommand{\bdm}{\begin{displaymath}}
\newcommand{\edm}{\end{displaymath}}
\newcommand{\beq}{\begin{equation}}
\newcommand{\eeq}{\end{equation}}
\newcommand{\bit}{\begin{itemize}}
\newcommand{\eit}{\end{itemize}}
\newcommand{\ben}{\begin{enumerate}}
\newcommand{\een}{\end{enumerate}}
\newcommand{\bfi}{\begin{figure}[htb]}
\newcommand{\bpfi}{\begin{figure}[p]}

\shorttitle{VLA-COSMOS Survey IV.}
\shortauthors{Schinnerer et al.}

\begin{document}


\title{The VLA-COSMOS Survey. IV. Deep Data and Joint Catalog}

\author{E. Schinnerer\altaffilmark{1},
M.T. Sargent\altaffilmark{1},
M. Bondi\altaffilmark{2},
V. Smol\v{c}i\'{c}\altaffilmark{3},
A. Datta\altaffilmark{4},
C.L. Carilli\altaffilmark{4},
F. Bertoldi\altaffilmark{3},
A. Blain\altaffilmark{5},
P. Ciliegi\altaffilmark{6},
A. Koekemoer\altaffilmark{7},
N.Z. Scoville\altaffilmark{5}
}

\altaffiltext{1}{Max-Planck-Institut f\"ur Astronomie, K\"onigstuhl 17,
    D-69117 Heidelberg, Germany}
\altaffiltext{2}{INAF-Istituto di Radioastronomia, Via Gobetti 101,
                 I-40129, Bologna, Italy}
\altaffiltext{3}{Argelander Institut f\"ur Astronomie, Universit\"at Bonn, 
                 Auf dem H\"ugel 71, D-53121 Bonn, Germany}
\altaffiltext{4}{National Radio Astronomy Observatory, P.O. Box O, Socorro, NM, 87801, U.S.A.}
\altaffiltext{5}{California Institute of Technology, MC 105-24, 1200 East
                 California Boulevard, Pasadena, CA 91125, U.S.A.}
\altaffiltext{6}{INAF-Osservatorio Astronomico di Bologna, Via Ranzani 1, 
                 I-40127 Bologna, Italy}
\altaffiltext{7}{Space Telescope Science Institute, 3700 San Martin Drive, 
                 Baltimore, MD 21218, U.S.A.}

\begin{abstract}
  In the context of the VLA-COSMOS Deep project additional VLA A array
  observations at 1.4\,GHz were obtained for the central degree of the
  COSMOS field and combined with the existing data from the VLA-COSMOS
  Large project. A newly constructed Deep mosaic with a resolution of
  2.5$''$ was used to search for sources down to 4$\sigma$ with
  1$\sigma \approx 12\,\mu$Jy/beam in the central 50$'\times$50$'$.
  This new catalog is combined with the catalog from the Large project
  (obtained at 1.5$''$$\times$1.4$''$ resolution) to construct a new Joint
  catalog. All sources listed in the new Joint catalog have peak flux
  densities of $\ge$5$\sigma$ at 1.5$''$ and/or 2.5$''$ resolution to
  account for the fact that a significant fraction of sources at these
  low flux levels are expected to be slighty resolved at 1.5$''$
  resolution. All properties listed in the Joint catalog such as peak flux
  density, integrated flux density and source size are determined in
  the 2.5$''$ resolution Deep image. In addition, the Joint catalog
  contains 43 newly identified multi-component sources.
\end{abstract}


\section{Introduction}

In recent years, several cosmological deep fields have been imaged at
20\,cm \citep[e.g.,
][]{ric00,bon03,con03,hop03,sey04,nor05,huy05,fom06,sim06,ivi07,sch07,mil08,owe08}
providing a few thousand radio sources down to flux limits of a few
10\,$\mu$Jy. These deep radio imaging data are sensitive enough to detect
star forming galaxies with star formation rates of several 10 to 100
$\solmy$ out to and beyond a redshift of $z$\,$\sim$\,1. Similarly, radio
galaxies can be seen out to redshifts of $z$\,$\sim$\,5 and the most
luminous ones even well into the epoch of reionization. Thus, deep
radio images in conjunction with deep imaging data at X-ray, optical
and infrared wavelengths are ideal to investigate the dust-unbiased
star formation, the evolution of radio(-loud) AGN, as well as the
population mix of radio sources in the first place.

In order to study the cosmological evolution of galaxies and black
holes, it is not only important to overcome the effect of cosmic
variance (e.g. by studying a large enough area) but also to understand
the effect of large scale structure on the evolution (e.g. by covering
a large contiguous area). To address the second effect in particular,
the Cosmic Evolution Survey (COSMOS)\footnote{~\tt
  http://cosmos.astro.caltech.edu} collaboration has conducted
  panchromatic imaging and spectroscopy of an equatorial field with a
size of 2\,$\rm deg^2$ \citep[for an overview, see ][]{sco07a} ranging
from X-ray XMM-Newton and Chandra \citep{has07,elv09}, UV GALEX
\citep{zam07}, optical and near-infrared ground-based
\citep{tan07,cap07a}, optical HST \citep{sco07b,koe07}, mid- to far-infrared
Spitzer \citep{san07}, millimeter \citep{ber07,sco08} and radio VLA
\citep{sch04,sch07} imaging to extensive optical spectroscopy
using the VLT/VIMOS and Magellan/IMACS instruments
\citep{lil07,tru07}. Most of these datasets are now publicly available
from the COSMOS archive at IPAC/IRSA\footnote{~\tt
  http://irsa.ipac.caltech.edu/Missions/cosmos.html}.

The VLA-COSMOS survey at 20\,cm is part of the overall imaging effort
and its scientific goals and motivation have been described in detail
by \cite{sch07}. Initial observations from a pilot project testing the
mosaicking strategy and giving a first source catalog are presented by
\cite{sch04}. As a large NRAO/VLA program, the VLA was used in A and C
configuration to cover the entire COSMOS field resulting in an image
with uniform noise properties in the central 1$\times$1\,deg$^2$ and
an average rms of 10.5\,$\mu$Jy. \cite{sch07} provide a detailed
description of the survey set-up, the data reduction, as well as the
testing and construction of the final VLA-COSMOS Large project catalog
(hereafter: Large catalog). Subsequently, \cite{bon08} derived the
completeness of the Large catalog and also analyzed the effect of
bandwidth smearing on the derived source flux densities to obtain the
source counts. Although the VLA-COSMOS Large Project dataset has been
used for several scientific results on, e.g. the faint radio
population, the radio-derived star formation rate density, the radio
AGN population and stacking of high-z galaxy populations
\citep{smo08,smo09a,smo09b,car07,car08}, the need for deeper radio
imaging data became apparent during the search for radio counterparts
to millimeter sources from the COSMOS MAMBO mapping data
\citep{ber07}; the Large project only provided counterparts for about
half of the mm-sources. Thus, the Deep project was initiated with the
aim of doubling the integration time for the central seven pointings
which fully cover the MAMBO 1.2\,mm map of the COSMOS field. These new
observations are described here.

The paper is organized as follows: After a description of the new
observations and the data reduction (\S \ref{sec:observations}), the
revision of the Large catalog (leading to a new version of v2.0) and
the construction of the new Deep catalog are outlined in \S
\ref{sec:newLarge} and \S
\ref{sec:deepcat}. In \S \ref{sec:identify} we explain the
construction of the final Joint catalog which is described in detail
in \S \ref{sec:joint_cat} where we also present all the tests and
corrections involved. A summary is given in \S \ref{sec:summary}.

\section{Observations and data reduction}
\label{sec:observations}

The central seven pointings of the mosaic for the VLA-COSMOS Large
project \citep[pointing numbers: F07, F08, F11, F12, F13, F16, F17;
see][]{sch07} were observed using the VLA A configuration in February
and March 2006. The observations were executed on 11 days, typically
lasting for a total of six hours starting at 07hr LST (Local Sidereal
Time). The exact coordinates of the pointing centers are listed in
Tab. \ref{tab:pos}. We used the same spectral set-up and set of
calibrators as for the VLA-COSMOS Large project to allow for an easy
combination of the data from the two projects: the VLA standard L-band
continuum frequencies of 1.3649 and 1.4351\,GHz in multi-channel
continuum mode, the quasar 0521+166 (3C\,138) for flux and bandpass
calibration at the beginning of each day, and the quasar 1024-008 as
phase calibrator. In order to obtain a good sampling of the uv-plane,
we cycled three times through the pointings each day with a total
integration time per pointing of $\sim$45 min. The final resulting
integration time per pointing was $\sim$8.25\,hr. In addition, we
changed the order of the observations of the individual pointings each
day to sample the uv plane uniformly in all pointings.

During the data reduction process, we excluded all EVLA
antennas\footnote{~A total of 3 EVLA antennas were included in the
  array at the time of the observations.} that were present in the
array during the observations. We followed the data reduction
procedure adopted for the Large project \citep{sch07} which uses
the Astronomical Imaging Processing System \citep[AIPS;][]{gre03}. uv
data points affected by RFI (radio frequency interference) were again
flagged by hand using the task \texttt{TVFLAG} and by excluding amplitudes
greater than 0.55 Jy. The flux density of the phase calibrator 1024-008
was found to be close to its previous value in the season of 2004/5.
After final calibration, the Deep data were combined with the uv data
of the corresponding pointings from the Large project.

\begin{figure}
\centering
\includegraphics[scale=0.9,angle=-90]{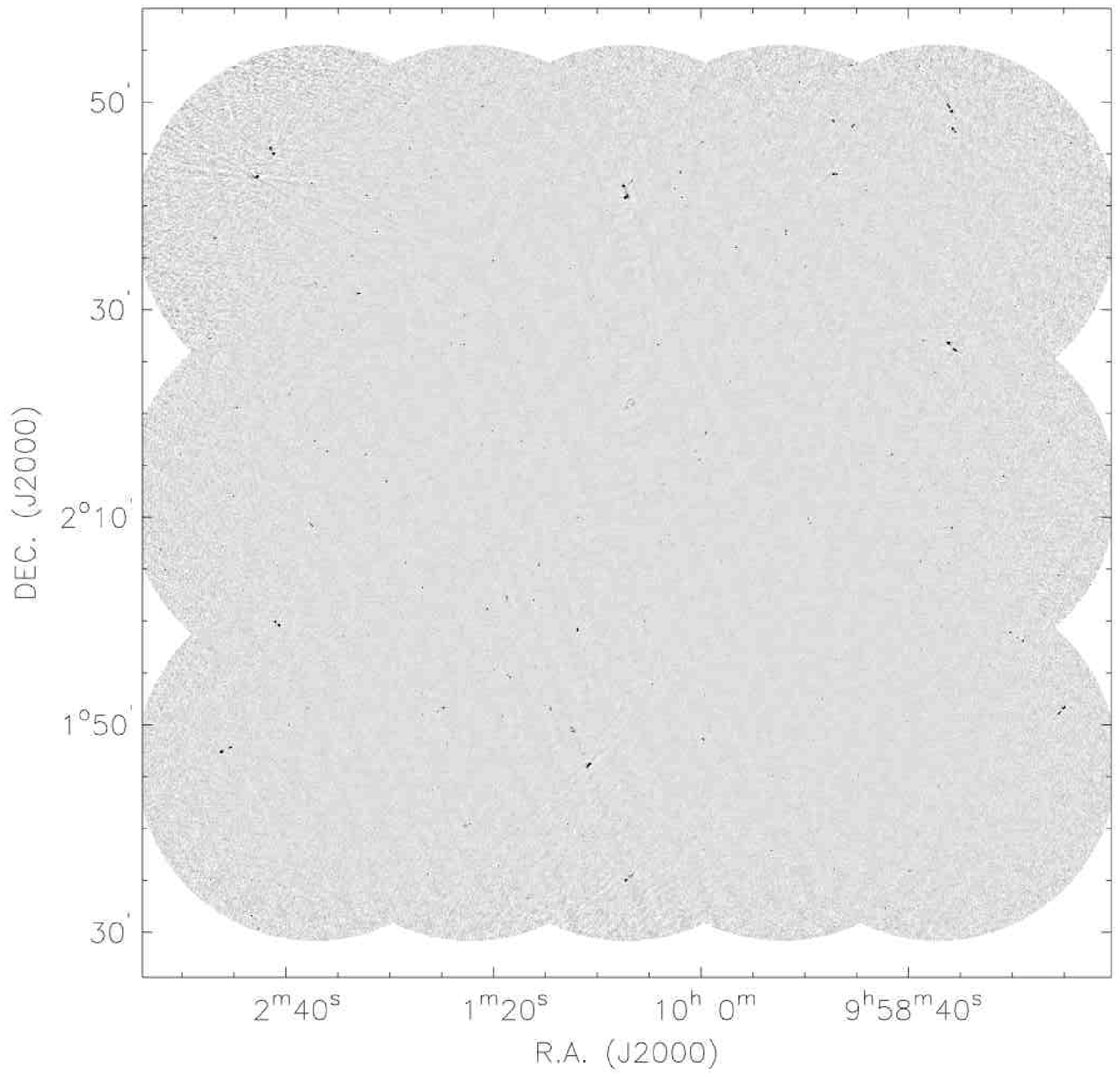}
\caption{The VLA-COSMOS Deep image at 2.5$\as$ resolution. \label{fig:deepimg}}
\end{figure}

\begin{figure}
\epsscale{.75}
\plotone{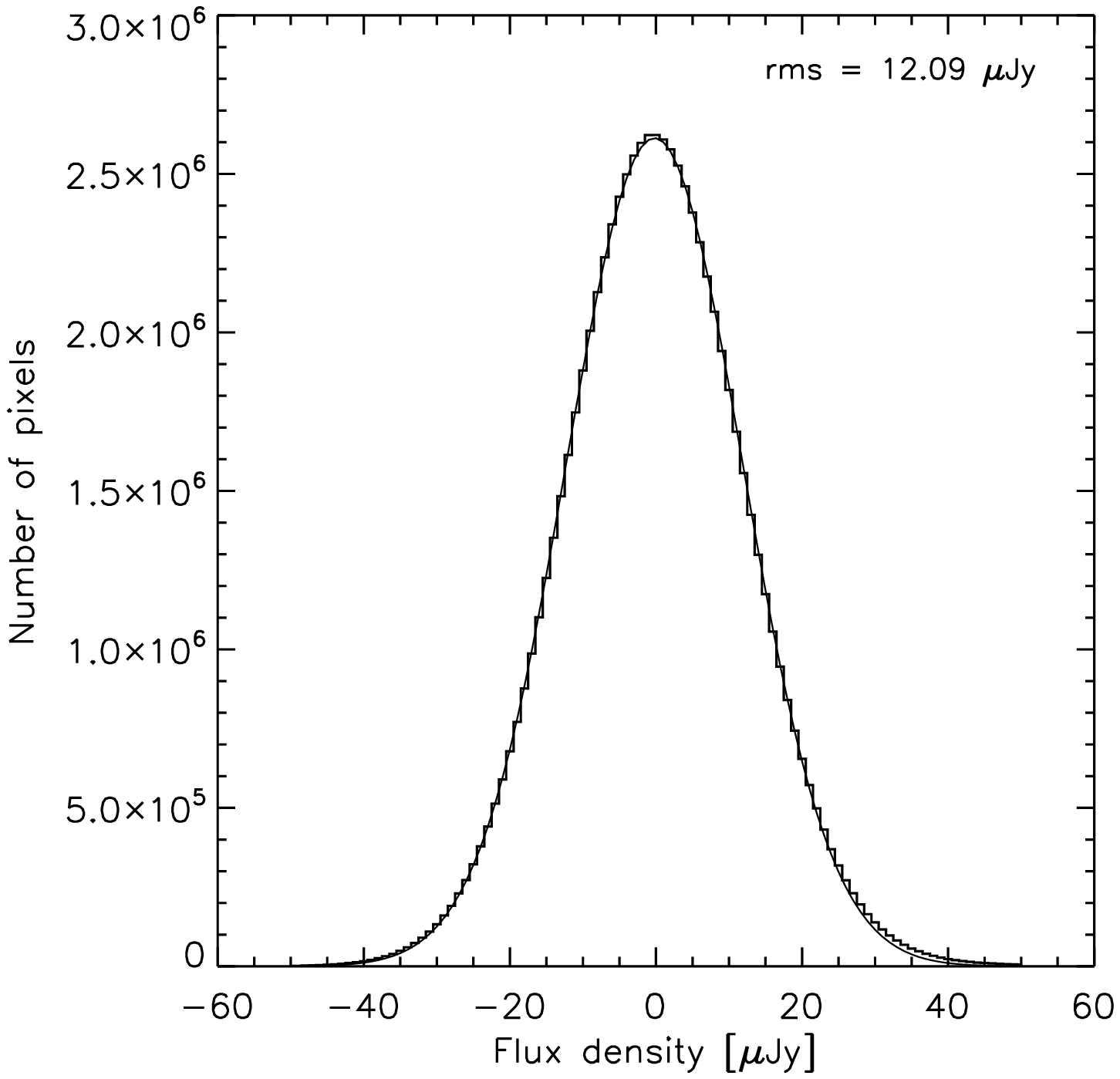}
\caption{Distribution of the noise in the VLA-COSMOS Deep mosaic.
  Pixel values extracted from a $52\am \times 52\am$ box centered on
  the COSMOS field center show a Gaussian distribution in agreement
  with our assumption of Gaussian noise. The fitted Gaussian (dashed
  line) has a rms of $\rm 12.09\,\mu Jy/beam$ ($\sigma$). The tail at
  high flux densities is consistent with the presence of
  sources in the image which were not excluded during the extraction
  process. \label{fig:rmshist}}
\end{figure}

\begin{figure}
\centering
\includegraphics[scale=.5,angle=-90]{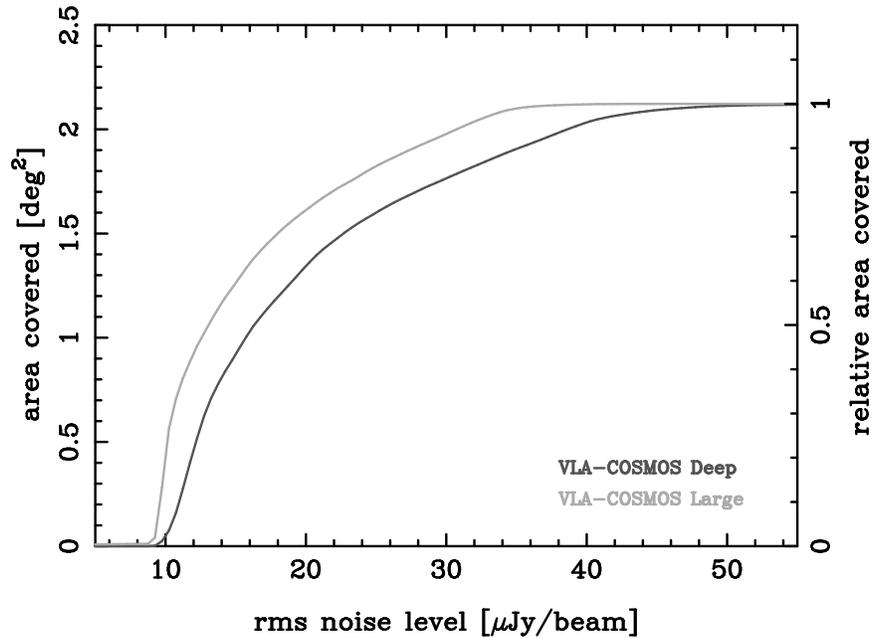} 
\caption{The rms level vs. the cumulative (left axis) and relative
  area (right axis) covered by the VLA-COSMOS Large (at a resolution
  of 1.5$''\times$1.4$''$) and Deep (with a resolution of
  2.5$''\times$2.5$''$) project mosaics.
  \label{fig:arearms}}
\end{figure}

The final imaging was performed using the task \texttt{IMAGR} with the
same settings as for the Large project, i.e. the same tiling of the
individual pointings and the associated \texttt{CLEAN} boxes were
used, the flux cut-off was kept at 45\,$\mu$Jy, and the restoring beam
was again set to 1.5$''\times$1.4$''$. The robust parameter was
changed from 0 to 0.25 to obtain a dirty beam as close as possible to
the final \texttt{CLEAN} beam. As for the Large project, each
polarization for each intermediate frequency (IF) was \texttt{CLEAN}ed
separately and combined afterwards in the image plane. The final seven
new fields were combined with the existing 16 fields from the Large
project to obtain a mosaic covering the full 2\,deg$^2$ of the COSMOS
field. The average rms achieved in the central 30$'$ was 8\,$\mu$Jy/beam
at the resolution of the Large project. In order to mitigate the
effect of bandwidth smearing on the derived peak flux densities during
source extraction (see \S \ref{subsec:bwsmeth}), the final mosaic was
convolved to a lower resolution of 2.5$''$ (Fig. \ref{fig:deepimg})
leading to an increase of the rms to $\sim$12\,$\mu$Jy/beam.  We tested
several convolution kernels by comparing the source fluxes and the
number of sources extracted for two representative sub-images of the
mosaic. The resolution of 2.5$''$ provided the best compromise between
minimizing the flux losses and maximizing the number of extractable
sources. We verified that the noise distribution in the Deep image
follows a Gaussian distribution as expected (Fig. \ref{fig:rmshist}).
This Deep mosaic was used for source detection and extraction (see \S
\ref{sec:deepcat}). Fig. \ref{fig:arearms} shows the fractional area
covered as a function of rms for the Deep and Large mosaics.

\section{The Revised Large Project Catalog}
\label{sec:newLarge}

The original source catalog for the VLA-COSMOS Large Project presented
by \cite{sch07} was created by the following procedure: 1) the AIPS
task \texttt{SAD} extracted candidate components down to a peak flux
density limit of 3$\sigma$, 2) after fits to the peaks of these
candidates using \texttt{MAXFIT} only those with an measured (rather
than the fitted Gaussian) peak flux density of $\ge$4.5$\sigma$ were kept,
3) components potentially missed by \texttt{SAD} were recovered with
\texttt{JMFIT} which was used to fit a Gaussian to all peaks above our
detection threshold that were not included in the \texttt{SAD}
component list, and finally 4) components not in the \texttt{SAD} list
but with non-zero major axes found in step 3 were added.  This
component catalog was transformed into a source catalog, after
multi-component sources, e.g.  large radio galaxies, were identified
by visual inspection.  Since in the case of faint sources, the
position and peak values of a Gaussian parametric fit might not be a good
representation of the real values, the position and value of the peak
found by \texttt{MAXFIT} replaced the results from the Gaussian fit.
Furthermore, all sources classified as unresolved (for our methodology
see Section 6.2 in \citealp{sch07}) had their integrated flux
density set equal to the peak flux density. The final VLA-COSMOS Large
project catalog (v1.0) has a total of 3601 entries.

Subsequent use of the Large catalog in conjunction with the COSMOS
optical and (near- to mid-)IR catalogs and images showed that the
fraction of spurious sources increases significantly below 5$\sigma$.
Spurious sources in our definition are sources that lie in a noisier
environment than expected for a $>4.5\sigma$ detection, i.e. three or
more 3$\sigma$ peaks are present in a 10$''\times$10$''$ box centered
on the source. The misidentification is likely due to a mismatch in
the derived rms value (in a 17.5$''\times$17.5$''$ box) and the local
rms distribution at the exact position of the source. This finding is
also confirmed by a comparison of sources present in the Large and
Deep catalogs (see \S \ref{subsec:singles}). Thus we exclude all sources
below 5$\sigma$ from the revised Large catalog. In addition, the
correction for bandwidth smearing as derived by \cite{bon08} has been
applied to the peak flux densities and the classification into
resolved and unresolved sources has been modified accordingly
\citep[for details see][]{bon08}. The revised version of the
VLA-COSMOS Large project catalog [hereafter: Large catalog (v2.0)]
contains a new total of 2417 sources. The revised VLA-COSMOS Large
project catalog (v2.0) has been publicly released to the COSMOS
archive at IPAC/IRSA\footnote{~\tt
  http://irsa.ipac.caltech.edu/data/COSMOS/tables/vla/}.

\section{Identification of Deep Project Sources}
\label{sec:deepcat}

\begin{figure}
\epsscale{1.}
\plotone{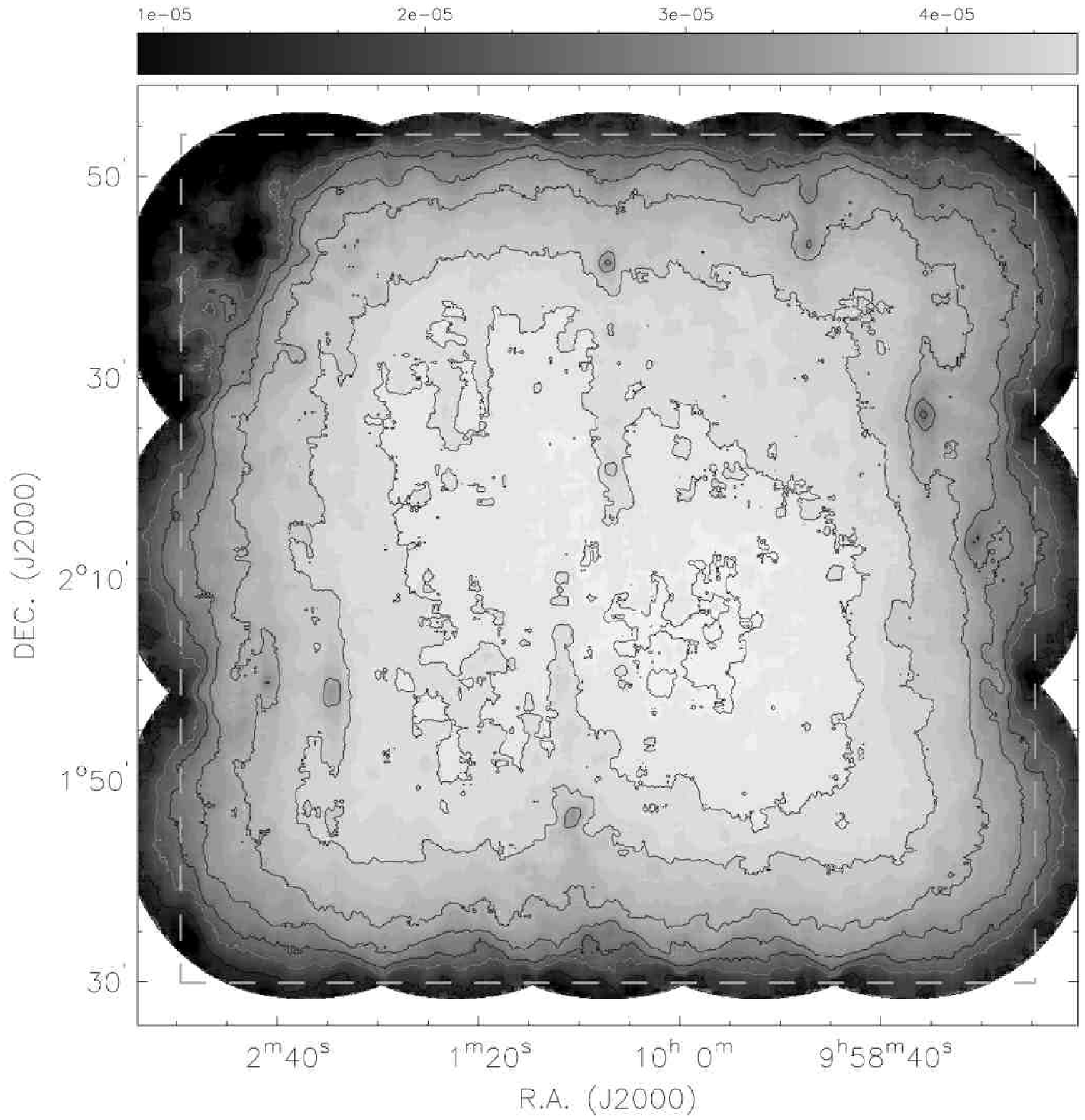}
\caption{Sensitivity map of the area covered by the VLA Deep Project,
  derived using the \texttt{AIPS} task \texttt{RMSD} (see text for details). 
  The rms is fairly uniform except for areas around strong radio sources. 
  Lighter shades indicate lower rms noise values. The contours correspond to
  rms levels of 10, 12, 15, 20, 25, 30, 34 and 40\,$\mu$Jy/beam. The search for
  radio sources was limited to the area enclosed by the (light grey)
  34\,$\mu$Jy/beam contour. For comparison, the dashed box indicates
  the area used to construct the catalog of the Large Project.
\label{fig:rmsimg}}
\end{figure}

As the rms is less uniform in the Deep mosaic than in the Large mosaic
and shows a steep increase towards the edges, applying a simple flux
density cut for source detection is not possible. In order to use the
AIPS source/component finding task \texttt{SAD}, a $S/N$ map was
created. First, a sensitivity map (Fig. \ref{fig:rmsimg}) was
constructed using the AIPS task \texttt{RMSD} with a box size of
105$''\times$105$''$ to estimate the rms and a sampling step size of
2.45$''$ in both RA and DEC. To obtain the final input $S/N$ map for
\texttt{SAD}, the Deep image was divided by the sensitivity map and
blanked at the edges where the rms values in the sensitivity map
exceeded 34\,$\mu$Jy/beam (the largest rms value present in the Large
mosaic). This last step allowed us to search for sources in the
corners/edges of the mosaic, whereas the area corresponding to the
COSMOS field was searched for the construction of the Large catalog
\citep[see][]{sch07}.

\texttt{SAD} was set to search for sources in six iterations with
cut-off levels for the peak of 10, 8, 7, 6, 5, and 4$\sigma$. The
derived values in units of $S/N$ were converted to mJy/beam using the
corresponding rms values from the sensitivity map. As in the case of
the Large project sources with complex structures -- e.g. radio
galaxies -- were fitted by multiple Gaussian components, such that the
derived catalog contains components rather than real sources. In order
to identify these multi-component sources, we compared by eye the
location of the multi-component sources from the Large catalog (v2.0)
with the component list obtained for the Deep image to identify the
components belonging to the same source. During the comparison with
the Large catalog (v2.0), it was noticed that several slightly
extended sources were fitted by two separate components with a typical
separation of $\sim$1$''$ (hereafter: `twin' sources). In order to
determine the nature (simple extended Gaussian vs. truly non-Gaussian
geometry) of these 'twin' sources, a single Gauss component was fitted
to these sources using \texttt{JMFIT}. If the integrated flux density
of this single Gaussian fit was consistent within the formal errors
with the combined integrated flux densities of the two components
derived by \texttt{SAD}, we replaced the `twin' sources by a single
object.

\section{Construction of the Joint Source List}
\label{sec:identify}

Based on the size distribution of faint radio sources
\citep[e.g.][]{mux05,fom06,owe08,bon08} a significant fraction of
sources should be resolved at our flux limit and resolution.
\cite{owe08} showed that source extraction at different resolutions is
ideal to maximize the number of sources found. Therefore, the
combination of the VLA-COSMOS Large and Deep catalogs is ideal to
find a maximum number of sources for inclusion in a joint catalog.
In the following we describe the steps taken to identify such
radio sources in the VLA-COSMOS mosaics.

In order to construct a list of radio sources from the two separate
catalogs, we used the following process that is described in detail
below: First, sources present in both catalogs were included (\S
\ref{subsec:both}), then selected sources detected only in the Large Project
mosaic were added (\S \ref{subsubsec:large-only}), as were -- finally
-- a number of sources present only in the Deep catalog (\S
\ref{subsubsec:deep-only}). In order to minimize the inclusion of
spurious sources at the low significance end, the local rms was
re-examined at 1.5$''$ and 2.5$''$ resolution. All sources having a
final peak $S/N \ge 5$ at at least one resolution were included.

An overview of the number of sources selected from the Large (v2.0)
and Deep catalog for inclusion in the Joint catalog is given in Tab.
\ref{tab:vlacos}, together with a listing of some key properties.

\subsection{Estimate of the Local Background Noise}
\label{subsec:noise}

\begin{figure}
\epsscale{.80}
\plotone{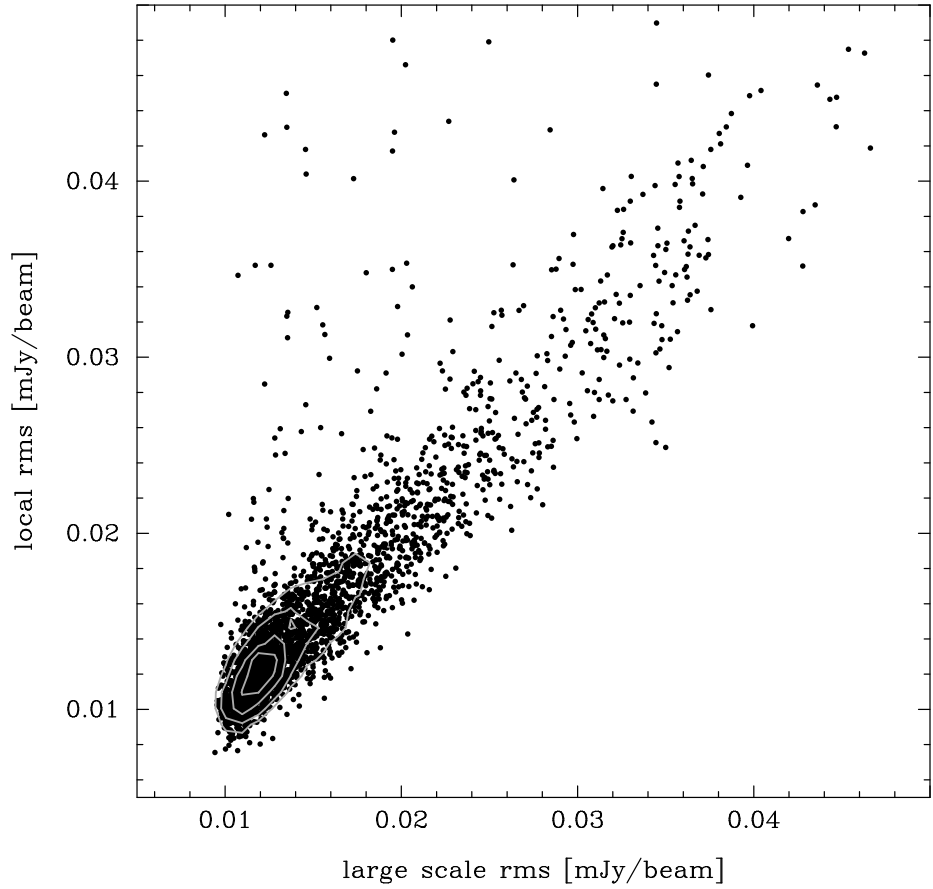}
\caption{Comparison of the noise estimate in the rms map used to generate 
the $S/N$ map for source detection in the 2.5$''$ resolution Deep image (original rms) and in 
the local rms map generated with a 50 pixel mesh size (corresponding to 17.5$''$). Contours
illustrate the distribution of points in the most densely populated part of the scatter plot.
  \label{fig:noise}}
\end{figure}

Based on the experience with the Large catalog (v1.0) it is important
to obtain an accurate estimate of the local rms to properly measure
the $S/N$ of individual radio sources. In order to find the optimal box
size for the local rms estimate, we compared the value in the rms map
used for source detection (see \S \ref{sec:deepcat}) to that obtained if
a smaller box is used. We chose a box size of
50 pixel (=17.5$''$) to obtain a good estimate of the rms around
compact sources. Fig. \ref{fig:noise} shows the different rms 
values obtained with a 105$''$ and 17.5$''$ box for
a number of low $S/N$ sources from the deep catalog and highlights the
importance of deriving an accurate local rms estimate. The choice was
based on the finding that a box size of 17.5$''$ used for the Large
catalog still resulted in a number of spurious sources at the low $S/N$
end and the requirement that at 2.5$''$ resolution the number of pixels
outside the source should still be sufficient\footnote{~At an image
resolution of $\theta_{\rm B}$ (FWHM), a 17.5$''$ rms box holds
\begin{equation}
N_{\rm beam} = \nicefrac{(17.5'')^2}{\Omega_{\rm beam}}\times \eta_h = \nicefrac{(17.5'')^2}{k_p\theta_{\rm B}^2}\times \eta_h
\end{equation}
independent, hexagonally packed beams. Here $\eta_h$\,$\approx$\,0.9
is the circle packing density \citep[e.g.,][]{wells81} and $k_p$\,$=$\,1.13 for
an aperture with a Gaussian power pattern \citep{ko64, otoshi81}. Hence,
as $N_{\rm beam}(2.5'')$\,$\approx$\,40, the 5\,$\sigma$ rms estimate has
an uncertainty of about 5\,$\sigma$/$\sqrt{40}$\,$\approx$\,0.8\,$\sigma$,
which implies that the local $S/N$ could be 4.2\,$\sigma$ or less in some
regions of the mosaic. To get an upper bound on how many of our sources
that were only detected with $S/N_{17.5''}$\,$\geq$\,5 at a resolution of 2.5$''$
might be affected by this, we checked which also had $S/N$\,$\geq$\,5 in
105'' rms boxes and found that 61\% also satisfied the latter criterion. These
were assigned a flag `det'\,=\,1 in the Joint catalog, while the remaining
sources are given `det'\,=\,2, see Section \ref{subsec:catdescript}
and Tab. \ref{tab:jointcat}.}.

We constructed new rms maps at 1.5$''$ and 2.5$''$ resolution using a box
size of 50 pixels with the task {\tt RMSD} and subsequently used them to
estimate the local rms. We refer to this
revised $S/N$ as the `local' $S/N$ ($S/N_{17.5''}$). All sources with a
$S/N_{17.5''} \ge 5$ will be included in the joint source list. 

\subsection{Sources Present in Both Catalogs}
\label{subsec:both}

For many sources in the Deep catalog it was possible to identify
counterparts in the Large catalog \citep{sch07} down to 5$\sigma$ or
even 4.5$\sigma$. The successfully matched sources include all but one of the objects
already known from the Large Project to consist of multiple Gaussian
components and the majority ($\sim$60\%) of the single component
sources in the Large catalog (v1.0).

\subsubsection{Identification of Single-Component Sources}
\label{subsec:singles}

To identify radio sources present in both the Deep and Large catalog,
we used the Large catalog down to 4.5$\sigma$. We matched the two
catalogs with a search radius of 1.5$\as$ to account for offsets
between the peak positions for extended objects. The spatial
characteristics of the noise in the Deep and Large image differ due
the additional coverage of the central seven pointings by the Deep
project and due to the differences in the final resolution (2.5$\as$
in the Deep mosaic vs. $\sim$1.5$\as$ in the Large mosaic; see \S
\ref{sec:observations}).

\begin{figure}
\epsscale{.8}
\plotone{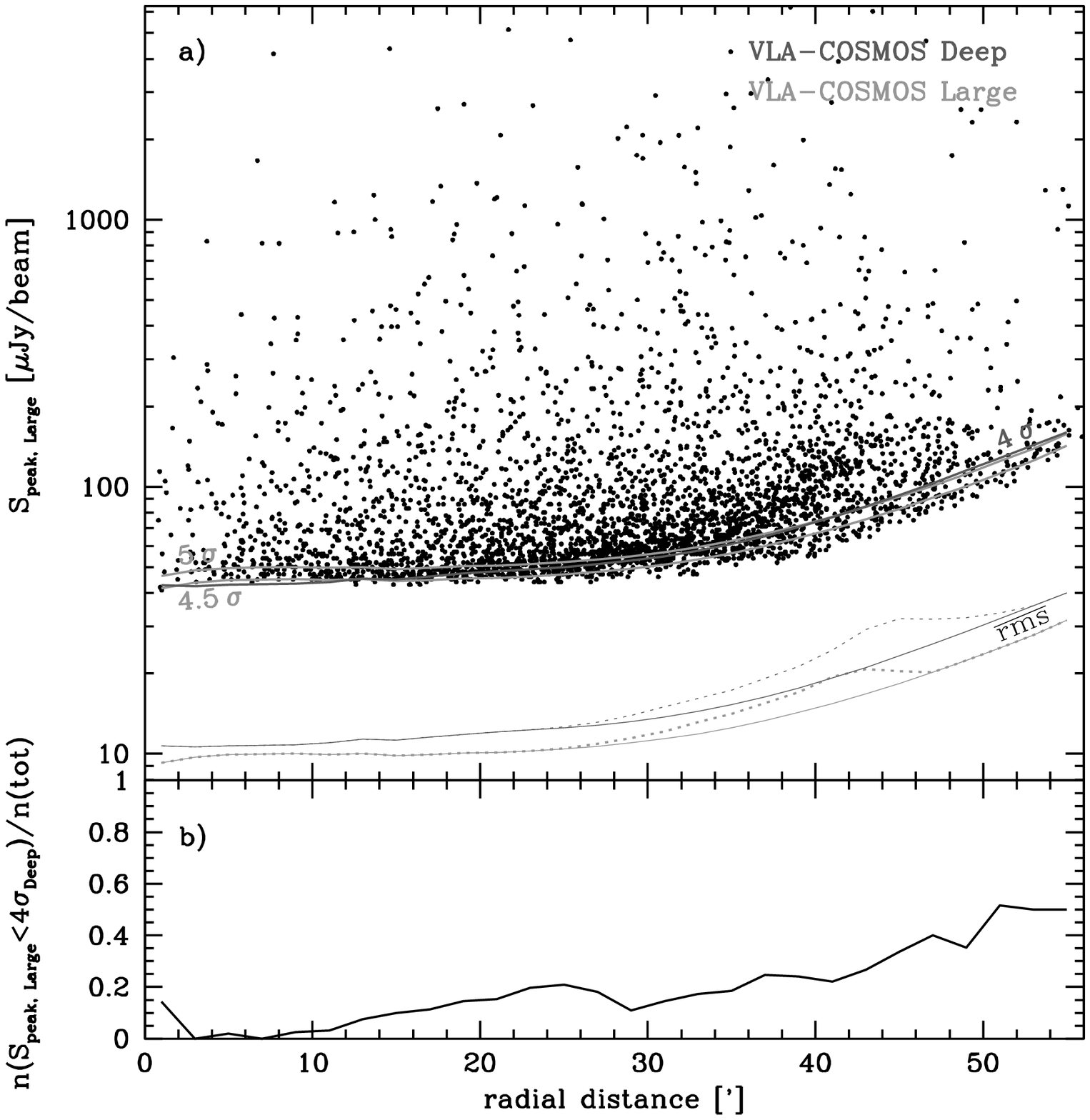} 
\caption{Correspondence of the detection thresholds used for the Deep
  and Large catalog. a) Peak flux densities from the Large catalog
  (v1.0) are compared to the mean rms noise level (${\rm \overline{rms}}$,
  averaged in concentric rings of width 2$\am$), and the detection
  thresholds applied in the VLA-COSMOS Large (light grey curves) and
  Deep mosaics (dark grey curve). A cubic spline fit to the measured ${\rm
    \overline{rms}}$ around the bump at $r \approx$ 45$\am$ gives an
  estimate of the minimal average noise level at a given radial
  distance from the center (solid lines at the bottom of the panel).
  b) Fraction of sources in the Large 
  catalog (v1.0) at a given radial distance $r$ that have a
  $S/N$ smaller than 4\,$\sigma_{\rm Deep}$ in the Deep
  mosaic.\label{fig:effsigma}}
\end{figure}

As mentioned in Section \ref{sec:newLarge} the occurrence of spurious
sources in the Large catalog is significantly reduced by omitting
sources in the range $S/N_{\rm Large}\in$ [4.5, 5[. The added depth
gained with the Deep project observations in the central 40$'$ (Fig.
\ref{fig:effsigma}) should increase the reliability of sources with
$S/N_{\rm Large}<$ 5 in the Large Project catalog (v1.0) within
this area. Fig. \ref{fig:effsigma} shows the corresponding detection
thresholds of the two images. The peak flux densities of sources in
the Large catalog (v1.0) are plotted as a function of their radial
distance from the field center, their lower envelope roughly tracing
the average $S/N$ detection threshold of 4.5$\sigma$ (used as cut
for the Large catalog (v1.0)). The radial evolution of the mean rms
(${\rm \overline{rms}}$) was calculated in concentric rings with a
width of 2$\am$ using the respective sensitivity maps for both images
(Fig. \ref{fig:rmsimg} for the Deep mosaic and Fig. 12 in \cite{sch07}
for the Large one). The `bump' in the measured mean noise level at a
radius $r\approx$ 45$\am$ is due to the sensitivity maps'
rectangular geometry. In order to obtain an estimate of the minimal
average noise level at this distance from the field centre we fitted a cubic spline to the measured ${\rm
  \overline{rms}}$ in this region. The resulting curve for the Large
mosaic, when scaled to 4.5$\sigma$, defines a lower envelope
to the measurements (ignoring detections in sites with locally lower
rms noise). The 4$\sigma$ limit of the Deep catalog corresponds to the
4.5$\sigma$ limit of the Large catalog for the inner radii with $r \le
15'$ while it increases to the 5$\sigma$ limit of the Large catalog at
larger radial distances.

\begin{figure}
\epsscale{.7}
\plotone{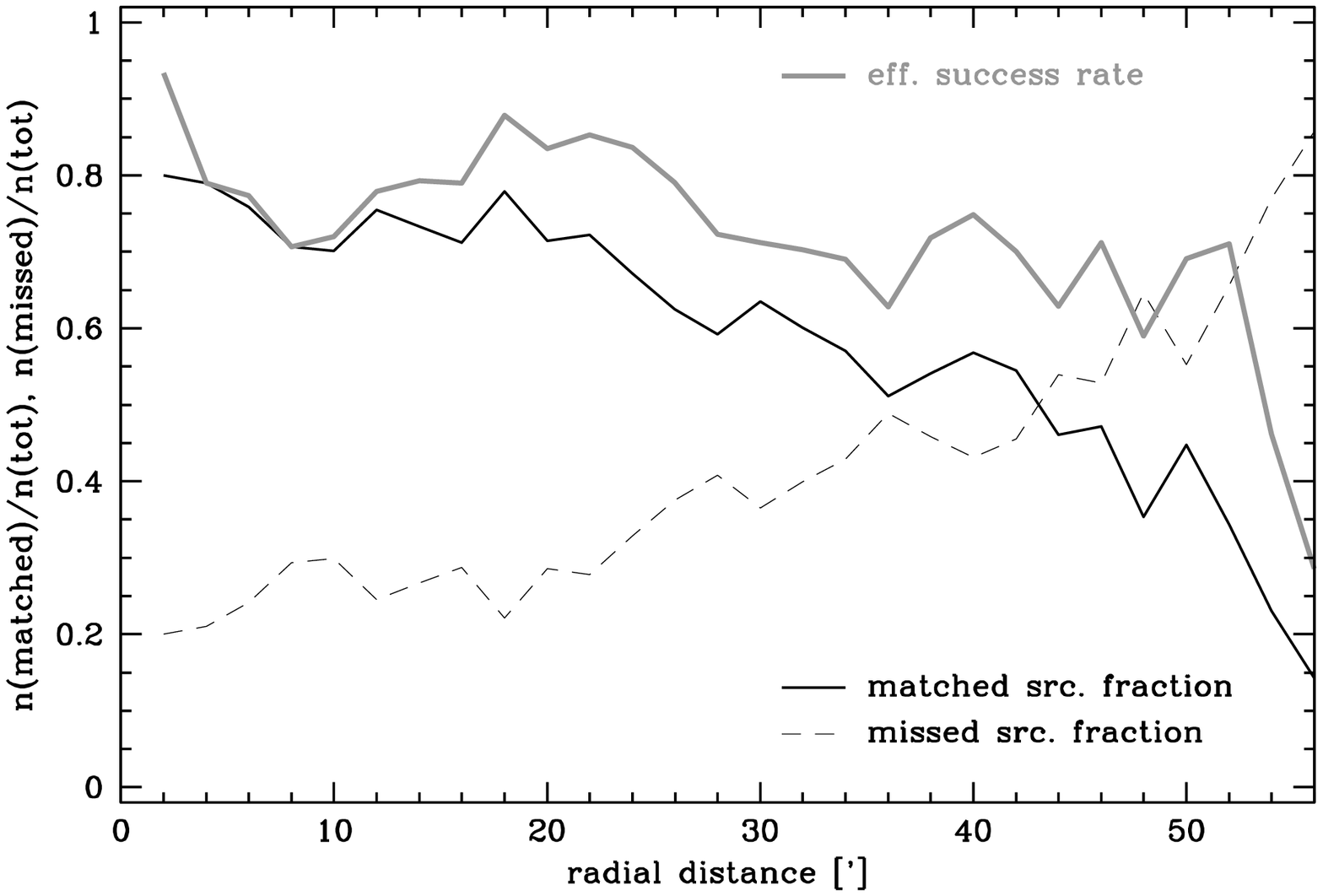} 
\caption{Fraction of sources in the VLA-COSMOS Large catalog (v1.0) which
  could (solid black line) or could not (dashed black line) be
  successfully assigned a counterpart within 1.5$\as$ in the
  VLA-COSMOS Deep catalog. The effective success rate (grey line) is
  determined by accounting for the fraction of sources that lie below
  4$\sigma$ in the Deep image (cf. panel (b) of Fig.
  \ref{fig:effsigma}.) and thus cannot be included in the Deep
  catalog. The vertical axis is normalized to the total
  number of sources in the catalog that lie in a given concentric
  ring of width 2$\am$. \label{fig:succrate}}
\end{figure}

The fraction of sources in the VLA-COSMOS Large catalog (v1.0) which
are at a given distance from the field centre and have $S_{\rm peak} <
4\sigma_{\rm Deep}$ is shown in Fig. \ref{fig:effsigma}b. The increase
of this fraction with distance from the field center will affect the
radial dependence of the number of successful matches between the two
catalogs. We verified that the success rate of the catalog matching
shows no distance dependence (Fig. \ref{fig:succrate}). While the
fraction of sources in the Large catalog (v1.0) for which a
counterpart could be identified decreases at larger distances (solid
black line), the effective matching success rate defined as ratio of
the measured matching success rate and the fraction of sources with
peak flux densities below 4$\sigma$ in the Deep mosaic stays basically
constant (thick grey line) within a range of 60 and 80\%.

Thus we conclude that the direct positional correlation of catalog
entries has not caused sources to be systematically missed as a
function of (radial) field distance in the VLA-COSMOS Deep mosaic. 

Before final inclusion in the source list we checked for each of the
successfully matched sources that it met our selection criterion of
peak $S/N_{17.5''} \ge 5$ at either 1.5$''$ or 2.5$''$ resolution in
the deep data. If this requirement was not satisfied the corresponding
source was not considered any further.

\subsubsection{Identification of New Multi-Component Sources}
\label{subsubsec:newmult}

As already stated above, all but one of the multi-component sources listed in the
Large catalog (v2.0) were also present in the Deep catalog. However,
during the merging of the two catalogs 43 new multi-component sources
were identified. Image cutouts of all new multi-component sources are
shown in Fig.  \ref{fig:multsrcimg}.

During the cross-correlation of the Large and Deep catalog several
sources were recognized to constitute sources made up of several
(Deep) rather than just a single (Large) Gaussian component during
visual inspection of all sources. In addition to this, several
Deep sources found in the neighborhood of multi-component objects
(identified in the Large catalog) were subsequently assigned to these
sources as additional components of their extended emission. In one
instance two of the former multiple component sources were joined to
form a new, larger multiple component source (a bipolar jet with
nucleus assigned the new Joint catalog ID
COSMOSVLADP\_J095758.04+015825.2: - the three merged components have
the following IDs in the VLA-COSMOS Large Project catalogs:
COSMOSVLA\_J095755.84+015804.2, J095758.04+015825.2 \&
J095800.79+015857.1). Besides these newly identified or augmented
multi-component sources there was also a small number of
multi-component sources that had either (i) not been previously listed
in the Large catalog or (ii) are situated outside the area searched during
the construction of the Large catalog.

\subsection{Sources present in only one catalog}
\label{subsec:singledetects}

Apart from the sources that are present in both catalogs, objects
present only in one of the two catalogs have been added to the list as
well. The exact procedures adopted are described in the following.

\subsubsection{Sources identified in the Large image only}
\label{subsubsec:large-only}

All sources in the Large catalog (v2.0) with a $S/N_{\rm Large} \geq
5$ but without a match in the Deep catalog are in principle valid
candidates for inclusion in the joint source list, provided they are not
flagged as detections potentially due to side lobes of strong radio
sources. Since \texttt{SAD} tries to fit Gaussians to all peaks above
a certain limit, the fact that certain sources were not found by
\texttt{SAD} during the construction of the Deep catalog suggests
that some of these objects might be mere noise peaks or have very
unusual morphologies. Similarly, we expect that a fraction of the
sources with $S/N<4.5$ might be real.

Therefore, we measured for all sources in the Large catalog (v1.0)
without a match in the Deep catalog the peak flux density in the Deep
image at 1.5$''$ and 2.5$''$ resolution at the according position from
the Large catalog following the steps described in \S
\ref{subsec:singcomp} and subsequently discarded 959 candidates with
$S/N_{17.5''} < 5$ at both resolutions while it was possible to
keep 393 of the unmatched sources with $S/N_{\rm Large} \geq 4.5$ from
the Large catalog (v1.0).

\subsubsection{Sources identified in the Deep image only}
\label{subsubsec:deep-only}

1155 objects -- predominantly in the range of 4$\le S/N_{\rm Deep}<$8
-- were detected in the Deep mosaic which did not have a counterpart
in the original Large catalog (v1.0). In order to assess the
significance of these sources which are potentially new detections due
to the different source extraction area for the Deep catalog, the
increased sensitivity in the central square degree and the fact that
the larger beam is more sensitive to slightly extended sources (as
\texttt{SAD} works with peak fluxes), we adopted the approach outlined
below. While the $S/N_{\rm Deep}=$ 4 cut corresponds at least to the
detection threshold used for the original Large catalog, our
experience with this catalog showed that the number of spurious
sources rises significantly below $S/N<$\,5 due to small mismatches
between the real local rms and the estimated value in the constructed
rms map.

Thus, the local $S/N$ was checked at 2.5$''$ resolution as well as at 1.5$''$
resolution. All sources with $S/N_{17.5''} \ge$ 5 were included in the
joint list. In total, 178 of 283 unmatched Deep detections
with $S/N_{\rm Deep}\geq$ 5 were taken over in the Joint catalog. Of 872
sources with 4 $\leq S/N_{\rm Deep}<$ 5 some 136 met our selection
criteria.

\section{Joint Catalog}
\label{sec:joint_cat}

In the following we describe how the flux densities and all
other measurements provided in the catalog were derived and which
corrections have been applied to the tabulated values.

\subsection{Correction for Bandwidth Smearing}
\label{subsec:bwsmeth}

Bandwidth smearing (BWS) can occur in radio synthesis imaging when a
finite frequency range is observed. It causes a radial smearing that
becomes more severe with increasing distance from the phase center (or
center of a pointing), as the phase calibration is mathematically
speaking only correct for a given frequency at the phase center. Thus
the effect is similar to chromatic aberration in optical imaging. A
detailed explanation and discussion of the effect of bandwidth
smearing for the VLA is given, e.g., by \cite{tho99} and \cite{bri99}.

For the VLA-COSMOS project with an observed bandwidth of 3.125\,MHz
for a single channel, a (final) radial smearing of 2.25$''$ is
expected at a radial distance of 15$'$ from the pointing center, i.e.
close to the cut-off radius of the individual pointings used when
creating the mosaic \citep[for equations and details,
see][]{bri99}. This bandwidth smearing effect also causes a decrease
of the peak flux density as the emission is now distributed over a
larger area. The integrated flux density, however, will not be
affected. As sources in the final mosaic can be covered by up to 7
individual pointings, the impact of bandwidth smearing strongly varies
at each location in the mosaic as a given source is separated by a
different amount from the centers of the individual pointings.

Thus, the effect of bandwidth smearing depends on the resolution
of the image. In correcting the measurements presented in the Large
catalog (v2.0) for the BWS effect \cite{bon08} adopted the approach of
boosting the measured values of the peak flux density $S_{\rm peak}$
by different amounts depending on the distance from the center of the
VLA-COSMOS field. The boost factors were derived from tests comparing
the peak flux densities of sources at the centers in individual
pointings and the derived flux densities in the final Large mosaic.
This approach can be used because the sensitivity map (cf. Fig. 12 in
\cite{sch07}) for the VLA-COSMOS Large project is fairly uniform over
significant areas, i.e. all locations suffer similarly from the
combined BWS effect of the contributing pointings. However, the
additional data for the central seven pointings for the Deep Project
changed the uniformity of the sensitivity distribution over the COSMOS
field. We thus examine both the method of \cite{bon08} as well as an
alternative based on a modeled sensitivity map for the VLA-COSMOS
Deep project. 

The model sensitivity distribution for an individual pointing was
constructed using the beam pattern for a VLA antenna at 1.465 GHz, as
specified in the help page for the \texttt{AIPS} task \texttt{PBCOR}:

\begin{equation}
1 - 1.343\times10^{-3}~\mathcal{X}^2 + 6.579\times10^{-7}~\mathcal{X}^4 - 1.186\times10^{-10}~\mathcal{X}^6~,
\label{equ1}
\end{equation}

\noindent 
where $\mathcal{X}$ is the product of the distance from the pointing
center in arcminutes with the observing frequency in GHz. The full
model sensitivity map (Fig. \ref{fig:modelsens}) was then assembled by
adding the individual beam patterns which were weighted according to
the observation time dedicated to each pointing (i.e., the central 7
pointings were weighted more strongly than the others by a factor of 1.4).

To estimate the magnitude of the required BWS correction we
closely follow the procedure described by \cite{bon08}. In brief, a
primary beam correction was applied to the images of the 23 individual
pointings and they were convolved to the same resolution of FWHM
2.5$\as$ as the Deep mosaic. Then we measured the peak flux densities
of selected sources in these individual images and compared their
values with those obtained from the same measurement carried out at
the corresponding position in the mosaic.

At this stage, it is mandatory to use sources with a peak flux density
that has not been significantly diminished by the BWS effect. The
dimensionless parameter $\beta=\frac{\Delta
  \nu}{\nu_0}\frac{\theta_0}{\theta_{\rm FWHM}}$ (see \cite{bri99} and
the VLA observational status
summary\footnote{~\texttt{http://www.vla.nrao.edu/astro/guides/vlas/current/node15.html}})
makes it possible to infer the amount by which peak flux densities are
reduced. In the case of the VLA-COSMOS Large image with an adopted
beam width of $\theta_{\rm FWHM}\sim$ 1.5$\as$ the peak flux densities
of sources within 5$\am$ of the pointing centers are expected to be
reduced by less than 5\%. At a resolution of $\theta_{\rm FWHM}\sim$
2.5$\as$, as in the case of the Deep image, the same small decrease
still exists as far as 8$\am$ from the pointing center. Thus we were
able to use a contiguous area covering a significant fraction of the
entire VLA-COSMOS mosaic (see area indicated by dashed white circles
in Fig. \ref{fig:modelsens}). A correspondingly larger number of
sources could thus be used to check the validity of the adopted method
for BWS correction. In the following we will quantify the magnitude of
the BWS effect using only sources within 5$\am$ of the center of
either of the 23 VLA-COSMOS pointings -- this ensures that peak flux
densities measured on individual pointings should differ by less than
2\% from the nominal value -- and we will use the larger number of
sources out to 8$\am$ to check the quality of the approach. Since the
aim is to correct for the BWS effect introduced by overlapping
pointings, strictly speaking no (valid) corrections can be derived for
sources located in the edges (i.e. those covered by only single
pointing, see Fig.  \ref{fig:bwscorrvar}).

Fig. \ref{fig:bwscorrvar} shows the comparison of the two different
methods to correct for the BWS effect. It should be noted that the
maximum decrease observed is only 10\%. The top row displays the
uncorrected values of the quantity $\mathcal{S} =S_{\rm peak,~mosaic}/S_{\rm peak,~pointing}$, i.e. the ratio of the
peak flux densities measured in the mosaic and in an individual
pointing, as a function of normalized sensitivity (left column) and
radial distance from the field center (right column). 
The median ratio
$\langle\mathcal{S}\rangle$ of peak flux densities
in a specific bin is marked by an open circle. The associated error
bars span the interquartile range of $\mathcal{S}$
within a bin in distance or model sensitivity. Median and errors are
reported in red when only sources within 5$\am$ of the center of one
of the 23 pointings are used (i.e.  only the red points) and in black
when all sources within 8$\am$ are considered (i.e. the red and grey
points).

It is obvious from the upper row of Fig. \ref{fig:bwscorrvar} that the
effect of bandwidth smearing disappears\footnote{~Note that at these
  radial distance the BWS effect can no longer be determined due to the
  lack of overlapping fields} (i.e. the value of the ratio
$\mathcal{S}$ approaches unity) towards the edge of
the field where there are no overlapping pointings or, conversely, at lower
sensitivity values which stem from regions in the field covered by
only one pointing. The best-fitting linear trends define which BWS
correction should be applied to sources lying at a given distance or
within a region of a specific sensitivity value, depending on the
method considered. In the middle row the ratios $\mathcal{S}_{\rm sensitivity~correction}$ are corrected according
to the second method using the model sensitivity map (Fig.
\ref{fig:modelsens}), i.e. the the slope of the line as a function of
sensitivity (upper left panel). Thus the distribution of $\mathcal{S}$ is
necessarily flat when plotted as a function of model sensitivity. This
flatness is also preserved when the corrected peak flux densities are
plotted against distance from the center of the field.  The BWS
correction (third row) based on the comparison of the peak flux
density ratios as a function of distance from the center of the
VLA-COSMOS field following \cite{bon08} also fares well in
straightening the distribution of corrected ratios $\mathcal{S}_{\rm distance~correction}$ with respect to both
distance (again, a requirement) and model sensitivity. However, a weak
correlation of peak flux density ratios with sensitivity is still
present after the application of this method. Therefore we conclude
that the sensitivity-based approach is slightly better for BWS
correction and will use it to compute the corrected peak flux density,
$S_{\rm peak,\,corr.}$, provided in the Joint catalog (see also Tab.
\ref{tab:jointcat}).

\subsection{Catalog Entries}

\subsubsection{Single-component sources}
\label{subsec:singcomp}

95\% of the sources in the VLA-COSMOS Joint catalog are well fit by a
single Gaussian component. The following five steps were required to
obtain the position, peak and integrated flux density, as well as the
size for each source. All measurements described in the
following were carried out on the Deep Mosaic. In particular,
this also applies to those sources which were not part of the initial
Deep catalog (see \S\ref{subsubsec:large-only}).

\begin{enumerate}

\item The (parametric) integrated flux $S_{\rm total}$ was derived
  using the task \texttt{SAD}. In the case of the `twin' sources,
  \texttt{JMFIT} was used instead (for details see \S
  \ref{sec:deepcat}). At the same time the convolved shape parameters
  were determined.

\item The peak flux density was obtained using the task
  \texttt{MAXFIT}\footnote{~Note that \texttt{MAXFIT} derives the
    maximum by fitting a quadratic function to a 3$\times$3 pixel map.
    Given that our Large (Deep) CLEAN beam is well sampled with at least 4 (7) pixels
    per axis the map value and the \texttt{MAXFIT} value agree within
    $\sim$0.2\%.}. The square search box for \texttt{MAXFIT} was
  centered on the position of the Gaussian fit from the previous step.
  We used an initial box size of 3 pixels that was increased in steps
  of 2 pixels if \texttt{MAXFIT} did not converge. The measured peak
  flux density $S_{\rm peak}$ was corrected for the effects of
  bandwidth smearing (see \S \ref{subsec:bwsmeth}) resulting in the
  corrected measured peak flux density $S_{\rm peak,\,corr.}$.

\item The source position was set to the position of the \texttt{MAXFIT}
  peak.

\item As a significant fraction of sources is resolved at the faint
  flux levels probed here, it is necessary to classify the sources
  into resolved and unresolved. We followed the approach used by
  \cite{sch07} and \cite{bon08}. The method relies on the fact that
  the ratio between integrated and peak flux density is a measure of
  the spatial extent of a radio source. A detailed description of the
  method is given in Appendix \ref{app} together with a description of
  a suite of Monte Carlo simulations which we use to calibrate our
  classification scheme. Fig. \ref{fig:resunres} is the diagnostic
  diagram used to identify the resolved sources based on our
  simulations. The line defining the lower boundary to the locus of
  resolved sources in Fig. \ref{fig:resunres} is given by the
  following equation\footnote{~Note that in previously published
    catalogs (see references above) using this approach, the line
    described by eq.  (\ref{equ2}) follows from logarithmically
    mirroring a lower envelope to the points below the line $S_{\rm
      total}/S_{\rm peak, corr.}$\,$=$\,1 above this very line. In
    Fig. \ref{fig:resunres} we show that the negative mirror image of
    eq. (\ref{equ2}) is a very conservative lower envelope of the kind
    just described. As a consequence, the classification into
    resolved/unresolved sources adopted in this paper based on the
    simulations in the Appendix, leads to a significantly larger
    fraction of sources being classified as `unresolved' in the Joint
    catalog.}:

\begin{equation}
S_{\rm total}/S_{\rm peak, corr.} = 0.35^{\nicefrac{-11}{(S/N)^{1.45}}}~,
\label{equ2}
\end{equation}

A discussion on possible systematic biases inherent to this method is
presented in the Appendix. We performed an independent check
of this classification method by adopting the {\tt JMFIT} approach
used by, e.g., \cite{mil08} and \cite{owe08}. As {\tt JMFIT} also
derives upper and lower limits on the deconvolved size of the major
and minor axis of a source, these limits can also be used to divide
sources into resolved and unresolved. If the upper limit of the major
axis was equal to zero, a source is considered unresolved. Comparison
between both methods shows that $<$1\% of our resolved sources would be
considered unresolved using the {\tt JMFIT} criterion while 66\% of
the unresolved sources would be resolved. Thus the total fraction of
resolved sources would increase to $\sim$70\% similar to the
fractions found by \cite{mil08} and \cite{owe08}. Using equation (\ref{equ2}) we have
found 405 resolved sources for which the integrated flux is given by
the total flux of the Gaussian \texttt{SAD}/\texttt{JMFIT} fit, and
2329 unresolved sources for which the integrated flux is \`a
posteriori set equal to the corrected peak flux density. These numbers
correspond to 14\% and 81\% of the sources in the Joint catalog,
respectively, while the remaining 5\% represent radio sources with
multiple Gaussian components (see following section).

\item The deconvolved source size parameters (major axis $\rm
  \theta_{M,\,dec}$, minor axis $\rm \theta_{m,\,dec}$ and position angle
  $\rm PA_{dec}$) were calculated for the resolved sources according
  to the algorithms of the \texttt{AIPS} task \texttt{JMFIT}. For
  unresolved sources all three are set to zero.

\end{enumerate}

\subsubsection{Multi-Component Sources}

For sources consisting of multiple Gaussian components the integrated
flux density was manually measured on the VLA-COSMOS Deep image with
the \texttt{AIPS} task \texttt{TVSTAT} in order to encompass all the
emission from these irregularly shaped sources. As the individual
components all have flux density peaks of their own, no peak value is
specified for multi-component sources in Tab. \ref{tab:jointcat}.
Positions were adopted from the VLA-COSMOS Large project catalogs
whenever possible. In all other cases the position was chosen on an
individual basis based on the radio morphology of the sources. In
practice this usually corresponds to a luminosity weighted mean of the
positions of the subcomponents \citep[as already done for the
multi-component sources in the Large Project catalog, see \S 6
of][]{sch07}.

Measurements of the major and minor axes along the direction of
maximal extension of the source and at right angles to the latter,
respectively, are provided for multi-component sources. However, no
estimate of the position angle is provided as it is usually impossible
to define for the complex geometry of these sources. Note that while
the coordinates were taken over, fluxes and dimensions of
multi-component objects adopted from the Large catalog (v2.0) were
re-measured in the Deep project mosaic at 2.5$''$ resolution.

\subsection{Errors on the Catalog Entries}

Below we summarize how the errors on the source properties listed in
the VLA-COSMOS Joint catalog were computed. In the case of the
single-component objects analytic formulae from the literature are
adopted. For the multi-component objects which have a manually
measured integrated flux density $S_{\rm total}$, the accuracy is about
10\% based on the comparison of repeated flux measurements on the same
objects by different people. For multi-component objects no error
estimates are provided for the source shape parameters as these are
meant as a rough measure of the angular size of the source. For the
same reason we also do not provide values for the error on the major
and minor axes of the Gaussian single component sources.

The error on the peak flux density of single component sources is
defined as the local rms noise at the position of the source.  For
unresolved sources the integrated flux density and peak flux density
are identical and hence the error on the integrated flux density also
corresponds to the local rms noise.

In calculating the error $\sigma_{S_{\rm total}}$ for resolved sources
in the VLA-COSMOS Joint catalog we follow the approach of \cite{hop03}
\citep[see also][]{bon03,sch04} based on the assumption that the
relative uncertainty $\sigma_{S_{\rm total}}/S_{\rm total}$ in the
integrated flux density is due to uncertainties $\mu_{\rm data}$ in
the data and uncertainties $\mu_{\rm fit}$ in the Gaussian fit:

\begin{equation}
\frac{\sigma_{S_{\rm total}}}{S_{\rm total}} = \sqrt{\left(\frac{\mu_{\rm data}}{S_{\rm total}}\right)^2+\left(\frac{\mu_{\rm fit}}{S_{\rm total}}\right)^2}~.
\label{equ3}
\end{equation}

\noindent The two factors beneath the square root in the equation are (see
\cite{win84} and \cite{con97}, as well as the explanations in
\cite{sch04}):

\begin{equation}
\frac{\mu_{\rm data}}{S_{\rm total}} = \sqrt{\left(\frac{S}{N}\right)^{-2}+ 0.01^2}
\label{equ4}
\end{equation}

\noindent (where $S/N=S_{\rm peak}/{\rm rms}$) and

\begin{equation}
\frac{\mu_{\rm fit}}{S_{\rm total}} = \sqrt{\frac{2}{\rho_S^2}+\left(\frac{\rm \theta_B \theta_b}{\rm \theta_M \theta_m} \right)\left(\frac{2}{\rho_{\rm M}^2} + \frac{2}{\rho_{\rm m}^2}\right)}~.
\label{equ5}
\end{equation}

\noindent Here ${\rm \theta_{B}}$ and ${\rm \theta_{b}}$ are the major and minor
axis of the beam (in the present case of a circular beam ${\rm
  \theta_{B}}={\rm \theta_{b}}=$ 2.5$\as$) and, analogously, ${\rm
  \theta_{M}}$ and ${\rm \theta_{m}}$ the major and minor axis of the 
measured (i.e. convolved) flux distribution. The $S/N$ values of the fit --
$\rho_S$, $\rho_{\rm M}$, and $\rho_{\rm m}$ -- are parameter-dependent:

\begin{equation}
\rho_X^2 = \frac{\rm \theta_M \theta_m}{\rm 4\theta_B \theta_b}\left(1+\frac{\rm \theta_B}{\rm \theta_M}\right)^{\alpha}\left(1+\frac{\rm \theta_b}{\rm \theta_m}\right)^{\beta}\left(\frac{S}{N}\right)^2~,
\label{equ6}
\end{equation}

\noindent 
with $\alpha=\beta=$ 1.5 for $\rho_S$, $\alpha=$ 2.5 and $\beta=$ 0.53
for $\rho_{\rm M}$, and $\alpha=$ 0.5 and $\beta=$ 2.5 for $\rho_{\rm
  m}$.

The errors on the position of single-component sources are

\begin{eqnarray*}
\Delta\alpha &=& \sqrt{\epsilon_{\alpha}^2 + \frac{\rm \theta_M^2}{4~{\rm ln(2)}\rho_S^2}~{\rm sin(PA)}^2 + \frac{\rm \theta_m^2}{4~{\rm ln(2)}\rho_S^2}~{\rm cos(PA)}^2}\\
\Delta\delta &=& \sqrt{\epsilon_{\delta}^2 + \frac{\rm \theta_M^2}{4~{\rm ln(2)}\rho_S^2}~{\rm cos(PA)}^2 + \frac{\rm \theta_m^2}{4~{\rm ln(2)}\rho_S^2}~{\rm sin(PA)}^2}~,
\end{eqnarray*}

where $\epsilon_{\alpha}=\epsilon_{\delta}={\rm \theta_B}/10 \approx
0.25''$ are the positional calibration errors on right ascension and
declination and PA is the measured position angle of the Gaussian fit.

\subsection{Description of the VLA-COSMOS  Joint Catalog}
\label{subsec:catdescript}

The information on the procedures and equations used to calculated
each entry in the VLA-COSMOS Joint catalog has been presented above.
For each source we list the source name, the ID of the source in the
VLA-COSMOS Large Project catalogs (where available), as well as the
derived source properties and the associated errors. Furthermore, the
source properties and flags associated with each object are explained.
All 2865 radio sources in the VLA-COSMOS Joint catalog are listed by
increasing right ascension in Tab. \ref{tab:jointcat} with the
following columns:
\\
\\
Column(1): Source name
\\[1ex]
Column(2): Source name in VLA-COSMOS Large Project catalog (set to
J999999.99+999999.9 if not present in the VLA-COSMOS Large catalog)
\\[1ex]
Column(3): Right ascension (J2000.0) in degrees
\\[1ex]
Column(4): Declination (J2000.0) in degrees
\\[1ex]
Column(5): Right ascension (J2000.0) in hexagesimal format
\\[1ex]
Column(6): Declination (J2000.0) in hexagesimal format
\\[1ex]
Column(7): rms uncertainty in right ascension in arcseconds
\\[1ex]
Column(8): rms uncertainty in declination in arcseconds
\\[1ex]
Column(9): Peak flux density and its rms uncertainty in mJy/beam
\\[1ex]
Column(10): Peak flux density corrected for bandwidth smearing in
mJy/beam
\\[1ex]
Column(11): Integrated flux density and its rms uncertainty in mJy
\\[1ex]
Column(12): rms measured in the \texttt{RMSD} sensitivity map in
mJy/beam
\\[1ex]
Column(13): Deconvolved major axis size ${\rm \theta_{M,\,dec}}$ in
arcseconds
\\[1ex]
Column(14): Deconvolved minor axis size ${\rm \theta_{m,\,dec}}$ in
arcseconds
\\[1ex]
Column(15): Deconvolved position angle of source ${\rm PA_{dec}}$
(counterclockwise from North) in degrees
\\[1ex]
Column(16): Flag for resolved and unresolved sources\footnote{~Note
  that the flag value of `-2' is only assigned to sources that were
  not in the VLA-COSMOS Large Project catalog. The flag `-1', on the
  other hand, is used for sources that were classified as resolved in
  the Large Project
  catalog but are listed as unresolved in the Joint catalog.}\\
\begin{tabular}{rcl}
-2 &--& unresolved only in VLA-COSMOS Deep image\\
-1 &--& resolved only in VLA-COSMOS Large image\\
 0 &--& unresolved in both VLA-COSMOS Large \&\\
 & & Deep images\\
 1 &--& resolved in both VLA-COSMOS Large \& Deep\\
 & & images\\
 2 &--& resolved only in VLA-COSMOS Deep image\\
\end{tabular}
\\[1ex]
Column(17): Flag for distinction of multi-component and single-component sources\\
\begin{tabular}{rcl}
0 &--& single component source\\
1 &--& multi-component source identified in VLA-COS-\\
& & MOS Large image\\
2 &--& multi-component source identified in VLA-COS-\\
& & MOS Deep image\\
\end{tabular}
\\[1ex]
Column(18): Flag for catalog membership\\
\begin{tabular}{rcl}
-1 &--& only detected in the VLA-COSMOS Large image\\
 0 &--& detected in both the VLA-COSMOS Large \& \\
 & & Deep image\\
 1 &--& only detected in the VLA-COSMOS Deep image\\
\end{tabular}
\\[1ex]
Column(19): Flag specifying at which resolution the source was detected with $S/N_{17.5''}\geq$ 5
\begin{tabular}{rcl}
-1 &--& detected with $S/N_{17.5''}\geq$ 5 only at a resolution\\
 & &  of 1.5$''$\\
 0 &--& detected with $S/N_{17.5''}\geq$ 5 at both 1.5$''$ and 2.5$''$ \\
 & & resolution\\
 1 &--& detected with $S/N\geq$ 5 only at a resolution of \\
 & & 2.5$''$, but in both the large and\\
 & & small scale (105$''$ and 17.5$''$ box size, respectively)\\
 & & rms map\\
 2 &--& detected only with $S/N_{17.5''}\geq$ 5 (but not in the \\
 & & large scale rms map) at a\\
 & & resolution of 2.5$''$\\
\end{tabular}

\subsection{Comparison between Large (v2.0) and Joint Catalog}

We compared the source properties in the flux-limited Large catalog
(v2.0) and in the Joint catalog for
those 2256 catalog entries which are common to both catalogs. The
different resolution in the VLA-COSMOS Deep and Large mosaics leads to
a different sensitivity to extended emission.  Hence, source shape
parameters cannot be expected to show exact agreement between the two
catalogs. Therefore the comparison is limited to the peak and
integrated flux densities. The results of the comparison are
illustrated in Fig. \ref{fig:peakcomp} and Fig. \ref{fig:totcomp}
for the peak and integrated flux density, respectively. In particular,
it is instructive to compare the peak values of sources that are
classified as unresolved in both catalogs.  These sources are
highlighted in light grey in Fig. \ref{fig:peakcomp} and are seen to
scatter well around the diagonal line of 1:1 correspondence between
the measurements from the Joint and Large mosaics.  The comparison of
the integrated flux densities (Fig. \ref{fig:totcomp}) shows that there is a
small population of objects (light grey points) lying above the
diagonal (i.e. their integrated flux density is higher in the Joint
catalog). These objects are sources that were previously classified as
unresolved but have acquired the status of resolved in the Joint
catalog, implying that their integrated flux densities have
necessarily increased when measured in the Deep mosaic. This class of
objects accounts for nearly all of the outliers in the plot.
Multi-component sources (dark grey crosses) in general lie also above
the diagonal bisector, reflecting the enhanced sensitivity to extended
emission in the Deep image. For most other objects the new and
previous estimate of integrated flux density is in good agreement and
hence also with flux density estimates of the NVSS and FIRST surveys
as shown in \cite{sch07}.

\section{Summary}
\label{sec:summary}

Continued analysis of the VLA-COSMOS catalog presented in \cite{sch07}
and the completion of the VLA-COSMOS Deep project motivated the
compilation of a new radio catalog for the COSMOS field. The
VLA-COSMOS Joint catalog was generated by combining the catalogs
of the VLA-COSMOS Large Project with a newly created source catalog
(Deep catalog) from the 2.5$''$ resolution Deep mosaic. This catalog is
already available for download by the public at the COSMOS archive at
IPAC/IRSA\footnote{~~\texttt{http://irsa.ipac.caltech.edu/data/COSMOS/tables/vla}}.
A comparison of the depth and areal coverage for a representative
sample of deep field radio surveys at 1.4\,GHz shows that the
VLA-COSMOS covers the largest area at its depth and angular
resolution (Fig. \ref{fig:radio_surveys}). Thus it should be well
suited to also study effects of the Large Scale Structure on the
presence/absence of radio emission.

The reduction and analysis of the deeper 20\,cm observations of the
central 7 pointings of the VLA-COSMOS projects using the VLA in A
configuration have been described in detail (also referred to as
VLA-COSMOS Deep project). In order to minimize the effect of bandwidth
smearing the Deep mosaic has a resolution of 2.5$''$ (compared to 1.5$''$
for the Large mosaic) and it was used to create a corresponding source
catalog using the task \texttt{SAD}. 

An input list for the new Joint catalog was compiled by combining the
revised Large catalog (v2.0) and the new Deep catalog. The criteria
were set such that no particular bias against slightly extended radio
sources was present when selecting the sources. All properties of the
radio sources listed in the Joint catalog have been derived in the
2.5$''$ resolution Deep mosaic.

The construction of the Joint catalog was motivated by the desire to
provide a catalog of bona-fide radio sources in the COSMOS field for
distinct science applications that are interested in the radio
properties of certain populations of galaxies. On the other hand the
revised Large catalog (v2.0) is flux-limited (in radio), has a fairly
uniform sensitivity coverage and its completeness is well
characterized \citep[see][]{bon08}, thus it is well suited for, e.g.,
studies of the faint radio population
\citep[such as, e.g.,][]{smo08,smo09a,smo09b}.

\appendix

\section{Bias on the Estimation of Integrated Fluxes and Source Sizes}
\label{app}

At low signal-to-noise, source fitting algorithms (e.g.
\texttt{JMFIT}) are increasingly liable to return false results, an
effect which is further amplified by the fact that there are multiple
free parameters (e.g. total flux, two angular size components) which
cannot be legitimately fixed to make the fit more robust. This
Appendix discusses biases in the {\it total} flux measurements in our
catalog which may arise as a consequence of this.  We will not study
the uncertainties, systematic or not, which might affect the angular
source sizes we quote in the catalog, as we consider these less
important for most users of the Joint catalog. Since the unresolved
sources in the catalog have a differently determined (\texttt{MAXFIT})
total flux, any biases in the Gaussian source fitting will only be of
importance for sources classified as resolved. How many sources are
affected is determined by the choice of classification scheme (see
Section \ref{subsec:singcomp}). Part of this section hence deals with
our choice of a criterion that both limits flux biases in the final
sample and also performs satisfactorily when it comes to separating
resolved from unresolved objects.

We first describe our simulations. We performed a set of Monte Carlo
simulations following the approach used for the Large project
\citep{bon08}. Circular mock sources with random flux densities down
to 30\,$\mu$Jy and sizes following the measured radio source counts
and an angular size distribution $\langle\theta\rangle \sim S^m$ with
an exponent m=0.5 \citep[see][for details]{bon08} were inserted into
the Deep mosaic.  They are not subject to bandwidth smearing effects.
We searched at the positions of these 16,000 sources using
\texttt{MAXFIT} followed by the application of \texttt{JMFIT} (single
component fit). We verified that this approach gives the same results
as using \texttt{SAD}.  Roughly 6,500 sources could be recovered with
a $S/N$\,$\ge$\,5. We also ran simulations with different source size
distributions which all returned qualitatively similar results. We
would like to remind the reader that the final numbers and errors
quoted below are sensitive to the adopted intrinsic angular size
distribution and therefore also to the resolution of the mosaic used.

As a significant fraction of sources is resolved at the faint flux
levels probed by the VLA-COSMOS project, it is necessary to classify
the sources into resolved and unresolved. We adopt the approach used
by \cite{sch07} and \cite{bon08} \citep[for applications to other
radio surveys, see][]{pra00,bon03}. The method relies on the fact that
the ratio between integrated and peak flux density is a measure of the
spatial extent of a radio source (given by the major and minor axis
$\theta_{\rm M}$ and $\theta_{\rm m}$) in comparison to the size of
the synthesized beam (with major and minor axis $\theta_{\rm B}$ and
$\theta_{\rm b}$):
\begin{equation}
S_{\rm total}/S_{\rm peak} = (\theta_{\rm M}~\theta_{\rm m})/(\theta_{\rm B}~\theta_{\rm b})~.
\end{equation}

On the other hand we can estimate the limiting (intrinsic) size
$\theta_{\rm limit}$ at a given $S/N$ above which a source could be
classified as resolved.  The threshold $\theta_{\rm limit}$ is
estimated using eq. (16) of \cite{con97}: according to this expression
the error in size ($\Delta\theta_{\rm src.}$) is proportional to $\rm
(S/N)^{-1}$ (see Fig.  \ref{fig:Condon_resunres}; dashed line) and a
point source would thus on average be liable to have a convolved size
$\theta_{\rm obs.}$\,$=$\, $\theta_{\rm B} \pm \Delta\theta_{\rm
  src.}$. All sources with intrinsic sizes below the lower (red) line
in Fig. \ref{fig:Condon_resunres}, i.e. $\theta_{\rm
  limit}$\,$=$\,$\sqrt{\theta_{\rm obs.}^2 - \theta_{\rm B}^2}$ -- the
inferred size of a point source subject to the $S/N$-dependent error
$\Delta\theta_{\rm src.}$, cannot be expected to be resolvable.

The ratio $S_{\rm total}/S_{\rm peak}$ of our `unresolvable' (black;
$\theta$\,$<$\,$\theta_{\rm limit}$) and `resolvable' sources (grey;
$\theta$\,$>$\,$\theta_{\rm limit}$) is plotted as a function of the $S/N$
with which the simulated sources are detected in Fig. \ref{fig:sim_resunres}.
After calibration with the simulated sources, this diagnostic diagram will
be used to identify the unresolved sources in the real catalog. By
necessity, measured values with $S_{\rm total}/S_{\rm peak}<1$ are
due to the influence of image noise on the determination of source flux
density and/or size. In general, the noise can both lead to an artificial
reduction or increase of the true flux density ratio (as peak and
integrated flux density are determined independently) implying that not
all sources with $S_{\rm total}/S_{\rm peak}>1$ are genuinely resolved
either. As can be seen in Fig. \ref{fig:sim_resunres} there is a fairly well-
defined locus above which only few unresolvable sources occur. We
approximate it by a line with functional form $S_{\rm total}/S_{\rm peak} =
a^{\nicefrac{-b}{(S/N)^c}}$. As our main goal is to minimize total flux
density biases, we define a rather conservative separating line:

\begin{equation}
S_{\rm total}/S_{\rm peak} = 0.35^{\nicefrac{-11}{(S/N)^{1.45}}}~.
\label{equ2app}
\end{equation}

This choice ensures that a minimal number of unresolvable mock sources
is classified as resolved (in the following a `resolved' source is
understood to lie above the line given by eq. (\ref{equ2app}) in Fig.
\ref{fig:sim_resunres}), especially at low $S/N$ where the errors on
total flux measurements from Gaussian fits are largest. A curve rising
more slowly toward low $S/N$ would have raised the number of
resolvable sources actually classified as resolved. It would also,
however,
have increased the number of misclassified unresolvable sources.\\
In the upper window of Fig. \ref{fig:sim_resunres} we illustrate which
fraction of resolvable (unresolvable) mock sources is classified as
unresolved (resolved) in a given range of $S/N$. In total, $\sim$38\%
of the resolvable and $\sim$3\% of the unresolvable sources are
`misclassified', leading to an overall success rate of nearly 87\%.

Given the classification of our mock sources into resolved and
unresolved following eq. (\ref{equ2app}), their total fluxes can now
be set to $S_{\rm total}$ \,$\doteq$\,$S_{\texttt{JMFIT}}$ and $S_{\rm
  total}$\,$\doteq$\, $S_{\rm peak}$, respectively (see Section
\ref{subsec:singcomp}), and any systematic flux biases at low $S/N$
quantified. In Fig. \ref{fig:comp_inout}, we show the difference
between the input ($St_{\rm in}$) and output ($St_{\rm out}$)
integrated flux densities as a function of $S/N$ (left-hand column)
and as histograms (right). As the mock sources were randomly injected
into the mosaic and no attempt has been made to avoid confusion with
other sources, the outliers in the distributions can be explained due
to confusion with real sources. However, the median derived from the
distribution should be unaffected by these outliers. The median
relative offset between flux values in different ranges of S/N is
listed in Tab. \ref{tab:comp_inout}.  In our lowest $S/N$ bin
(5\,$\le$\,$S/N$\,$\le$\,6) we detect a small bias of $\sim$5\% (in
the sense that recovered fluxes overestimate the true value) for all
sources which becomes even more negligible for higher significance
sources. Although this bias is signficantly larger for resolved
sources, its effect on statistical tools, such as source counts, is
minimal, as these use all sources and the fraction of resolved source
is at low S/N is very small. As the scatter in the derived properties
(at fixed $S/N$) is much larger, we conclude that a systematic
correction of the derived integrated flux on a source by source basis
is neither straightforward nor indispensable.

\acknowledgments The National Radio Astronomy Observatory (NRAO) is
operated by Associated Universities, Inc., under cooperative agreement
with the National Science Foundation. We would like to thank the NRAO
for their support during this project. We thank our referee Jim Condon
for his insightful comments for improving the paper. CC thanks the
Max-Planck-Gesellschaft and the Humboldt-Stiftung for support through
the Max-Planck-Forschungspreis. CC and AD acknowledge support through
NASA grant HST-GO-09822.33-A. MTS acknowledges support by the German
DFG under grant SCHI 536/3-2 and SCHI 536/3-3. VS acknowledges support
by the German DFG under grant SCHI 536/3-1. This project has been
(partially) funded by the DFG Priority Programme 1177 `Galaxy
Evolution'.

\bibliography{references}

\clearpage

\clearpage

\clearpage

\clearpage

\clearpage

\clearpage

\begin{figure}
\epsscale{1.}
\plotone{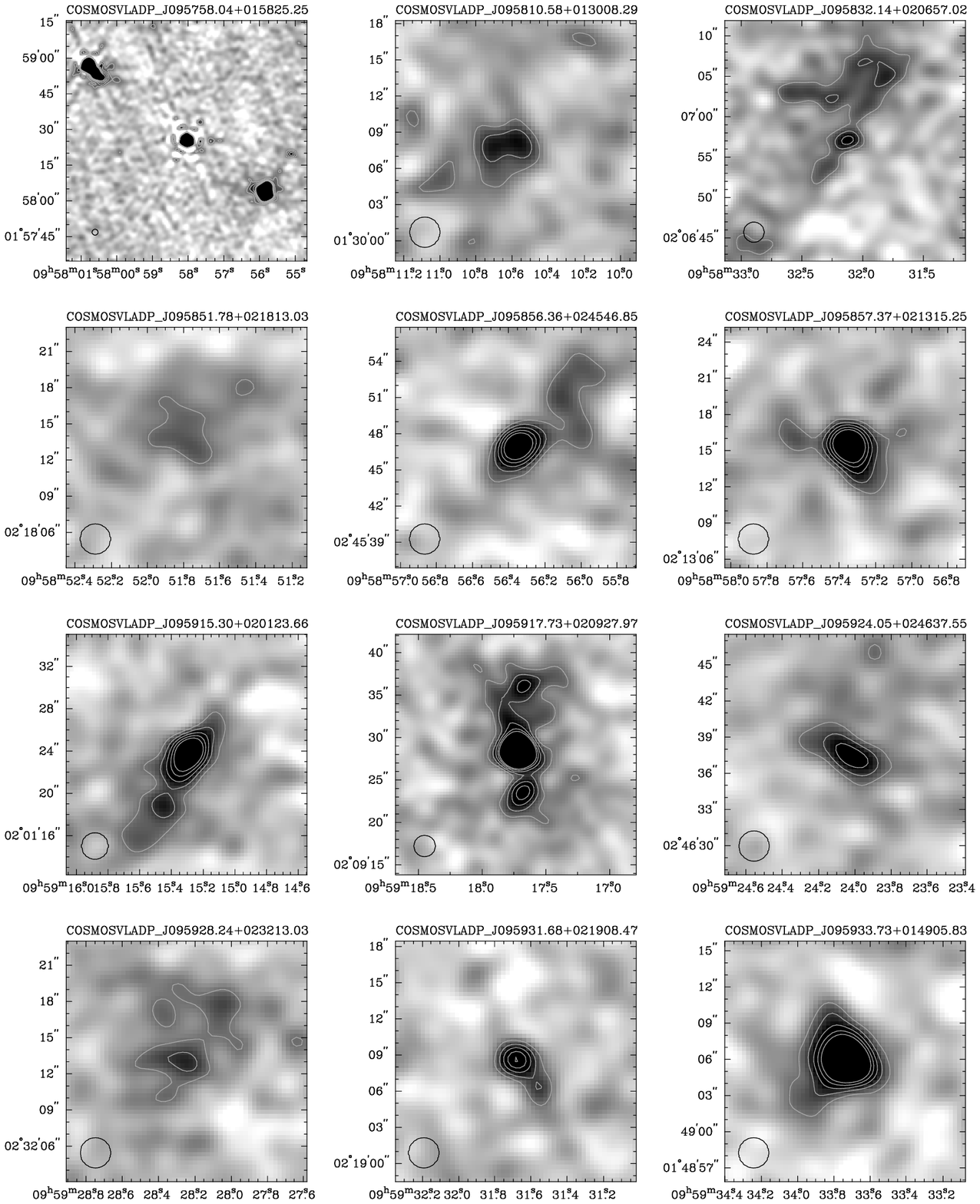} 
\caption{See last panel of figure for explanations.
  \label{fig:multsrcimg}}
\end{figure}

\addtocounter{figure}{-1}

\begin{figure}
\epsscale{1.}
\plotone{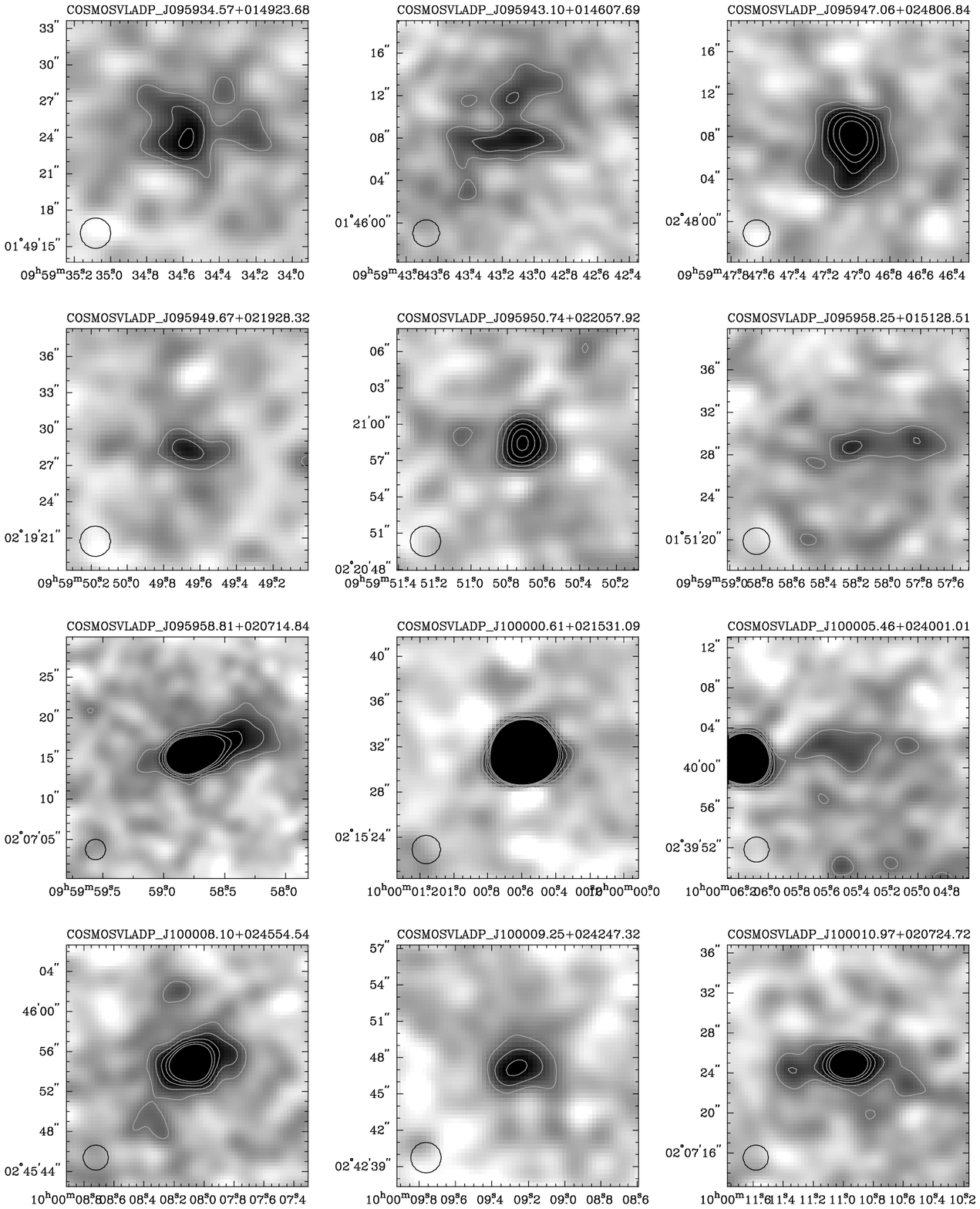} 
\caption{{\it cont.}}
\end{figure}

\addtocounter{figure}{-1}

\begin{figure}
\epsscale{1.}
\plotone{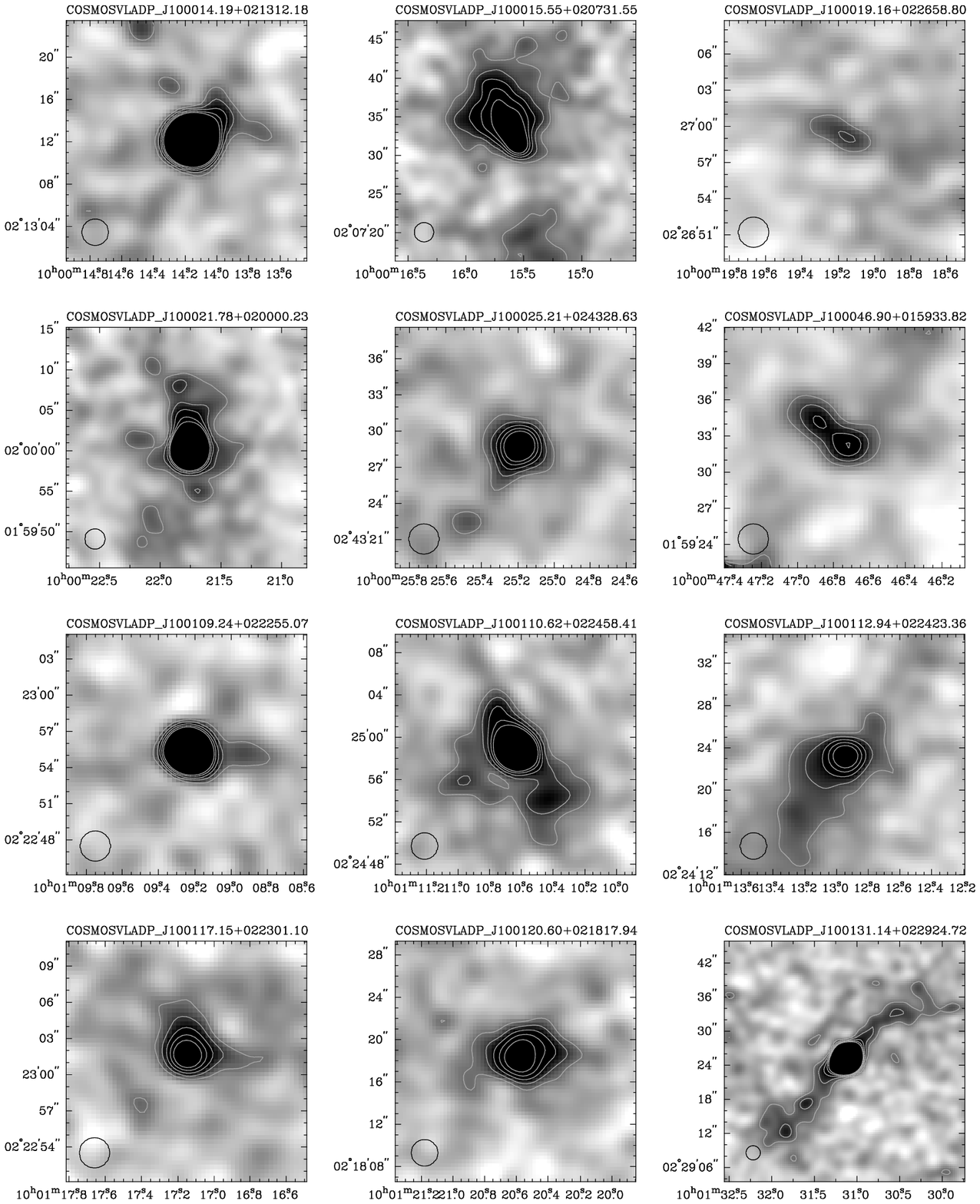} 
\caption{{\it cont.}}
\end{figure}

\addtocounter{figure}{-1}

\begin{figure}
\epsscale{1.}
\plotone{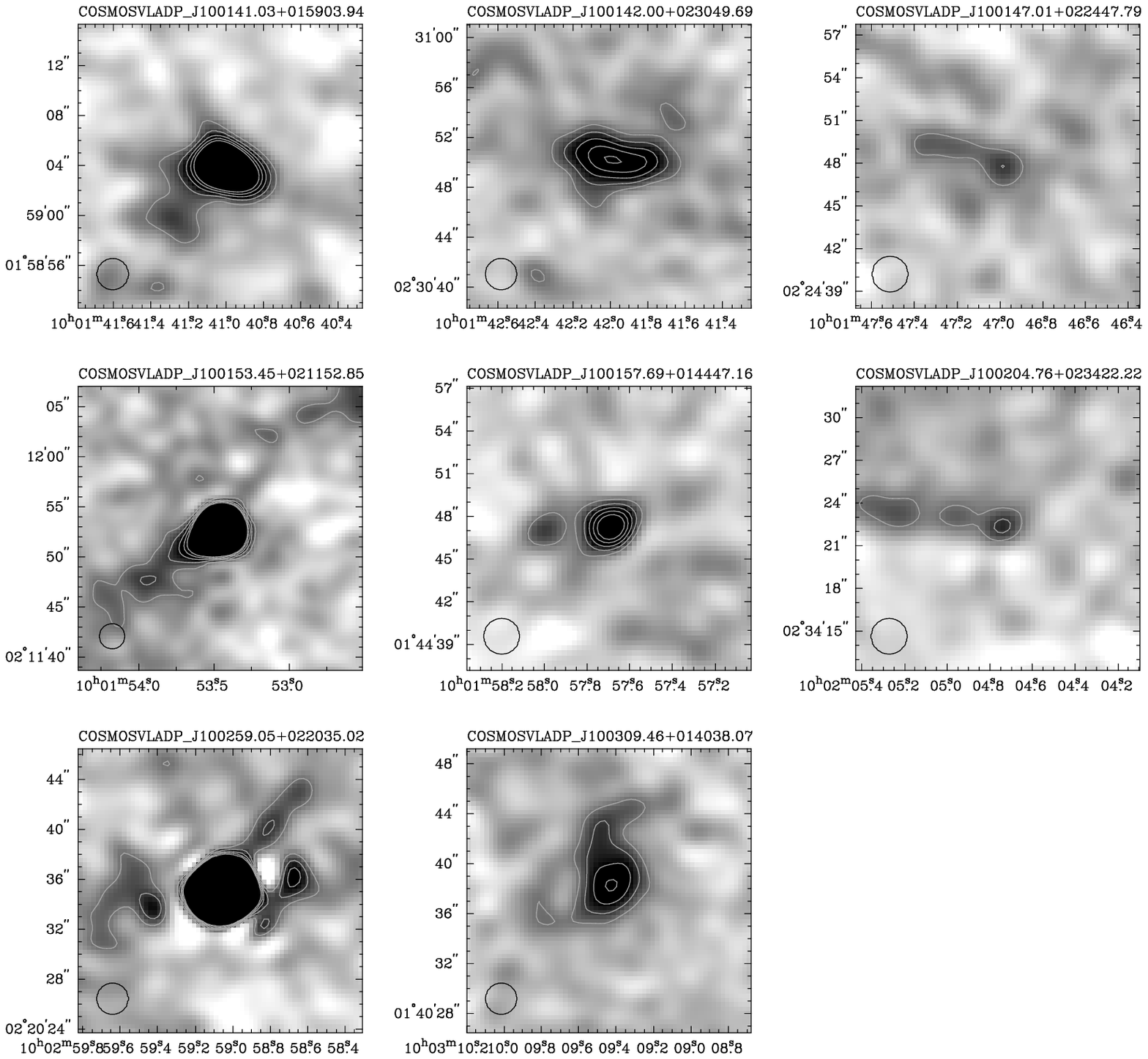} 
\caption{{\it cont.} Radio sources newly fitted with multiple Gaussian
  components in the VLA-COSMOS Deep image (cf. Fig. 16 of
  \cite{sch07} for images of all previously identified
  multi-component sources). The source name is specified at the top of
  the individual panels. The grey-scale ranges from -2$\sigma$ to
  7$\sigma$ of the local rms (Tab. \ref{tab:jointcat}). The contours
  start at 3$\sigma$ and increase to a maximum of 11$\sigma$ in steps
  of 2$\sigma$. The circular beam with FWHM of 2.5$\as$ is shown for
  reference in the lower left corner of each panel.}
\end{figure}

\clearpage

\begin{figure}
\epsscale{.74}
\plotone{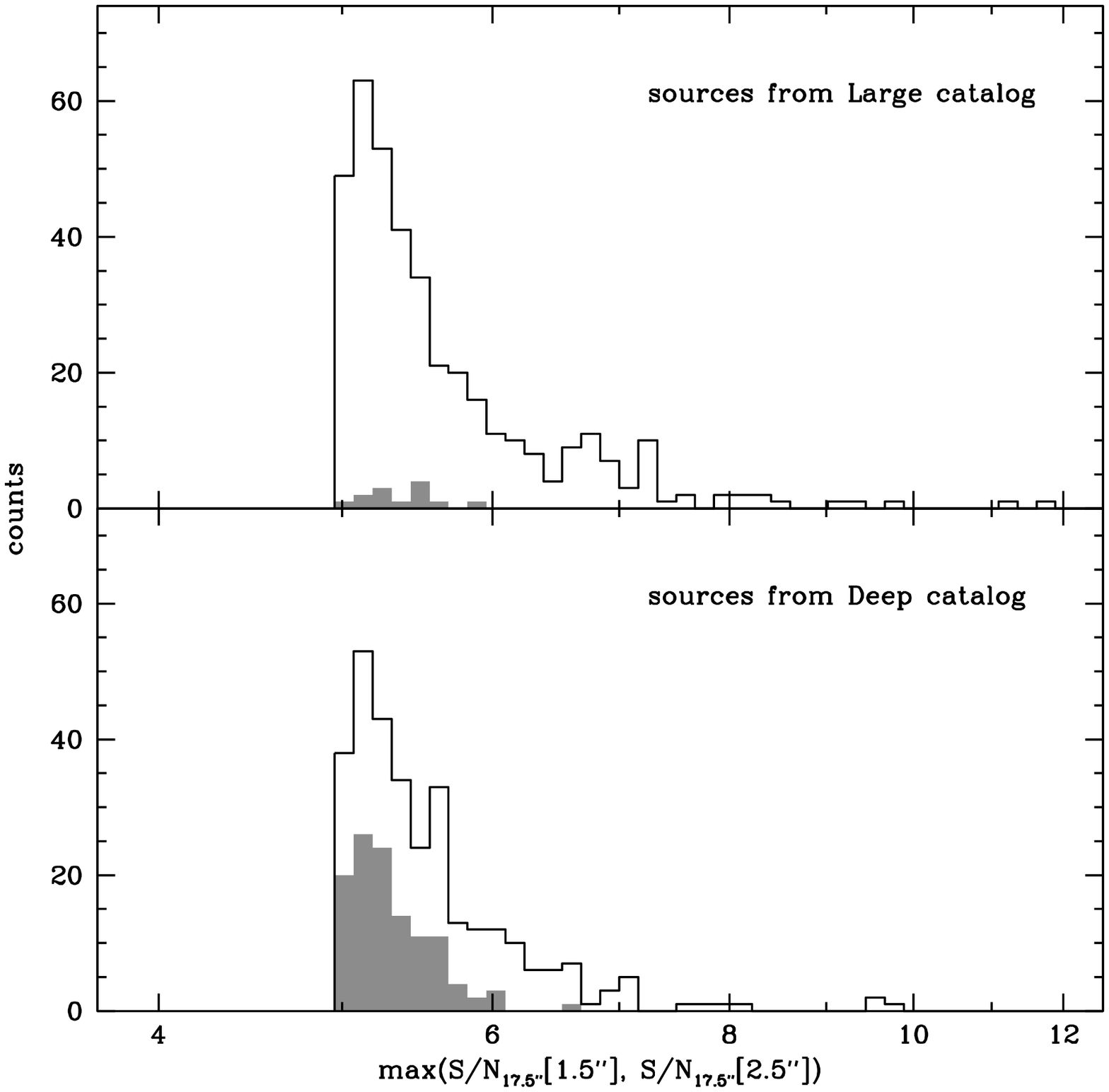} 
\caption{Distribution of the `local' $S/N$ measure, $S/N_{17.5''}$, for
  sources present only in the Large ({\it top}) or Deep ({\it bottom}) catalog.
  The highest measured $S/N_{17.5''}$ was adopted for each source (i.e. the
  larger of the two $S/N$ values determined at 1.5$''$ or 2.5$''$ resolution,
  $S/N_{17.5''}[1.5'']$ and $S/N_{17.5''}[2.5'']$, respectively). All
  sources with $S/N_{17.5''} \ge 5$ have been included in the Joint catalog.
  Grey histograms show the distribution of objects which are only detected
  at 2.5$''$ resolution and for which the detection threshold is only reached
  in the rms box of 17.5$''$ (flag `det' equals 2 in the Joint catalog; see Table
  \ref{tab:jointcat} and end of Section \ref{subsec:singles} for additional
  explanations), but not in the 105$''$ rms box used for the initial source
  extraction in the Deep mosaic. \label{fig:extras}}
\end{figure}

\begin{figure}
\epsscale{.74}
\plotone{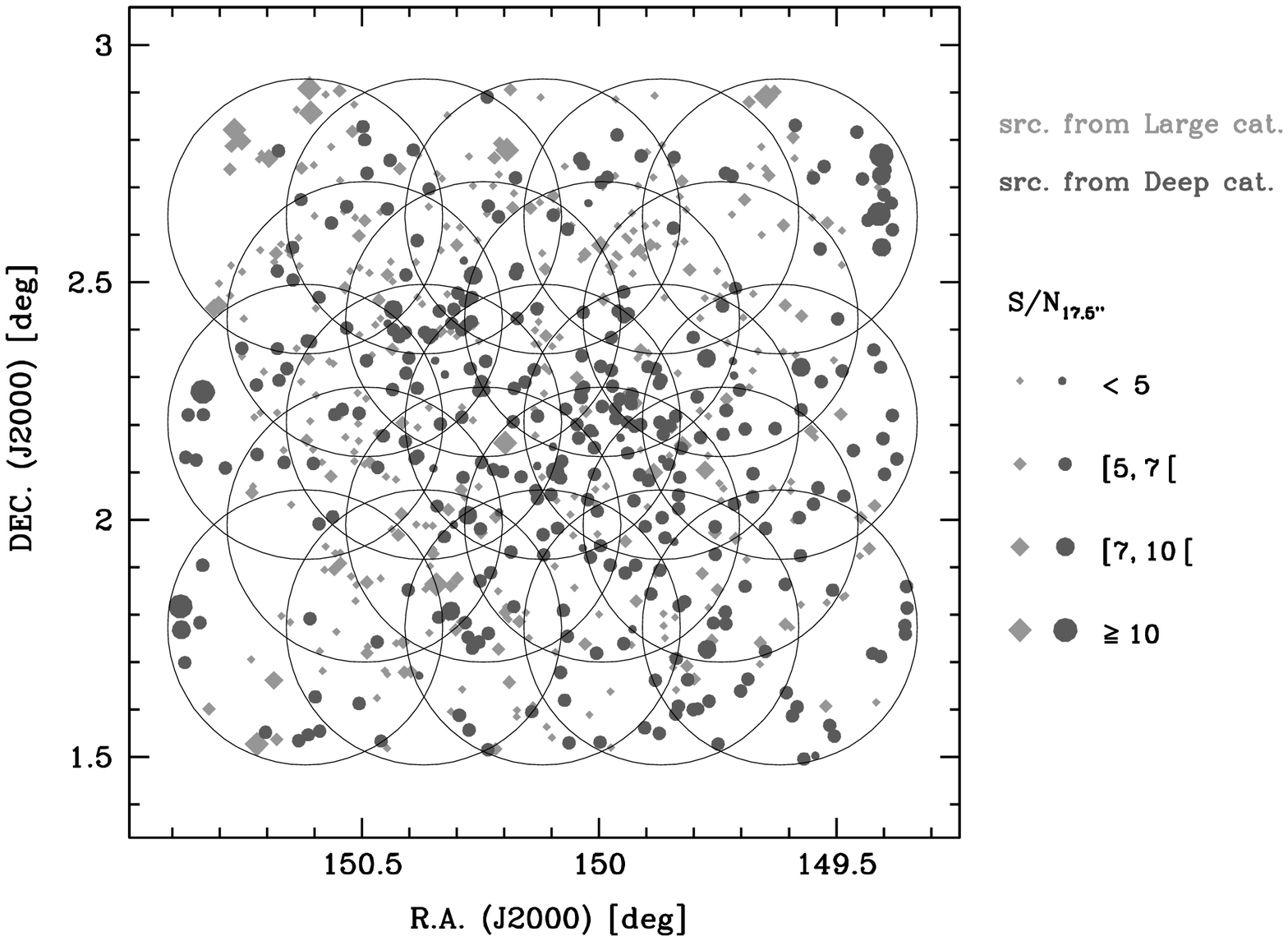} 
\caption{Spatial distribution of sources in the
  VLA-COSMOS Joint catalog that were either not listed 
  in the Deep (light grey) or Large catalog (v1.0) (dark grey). The
  size of the sign scales with the significance of the detection 
  (see scheme at right). Note that the newly identified bright sources
  along the eastern and western edge of the field did
  not figure in the VLA-COSMOS Large project catalog as the search
  area was previously restricted to the nominal size of the COSMOS field 
  (see Fig. \ref{fig:rmsimg}). 
  This geometric restriction was dropped in
  the detection of sources in the mosaic of the Deep Project.
  \label{fig:newnmissedpos}}
\end{figure}

\clearpage

\begin{figure}
\epsscale{1.}
\plotone{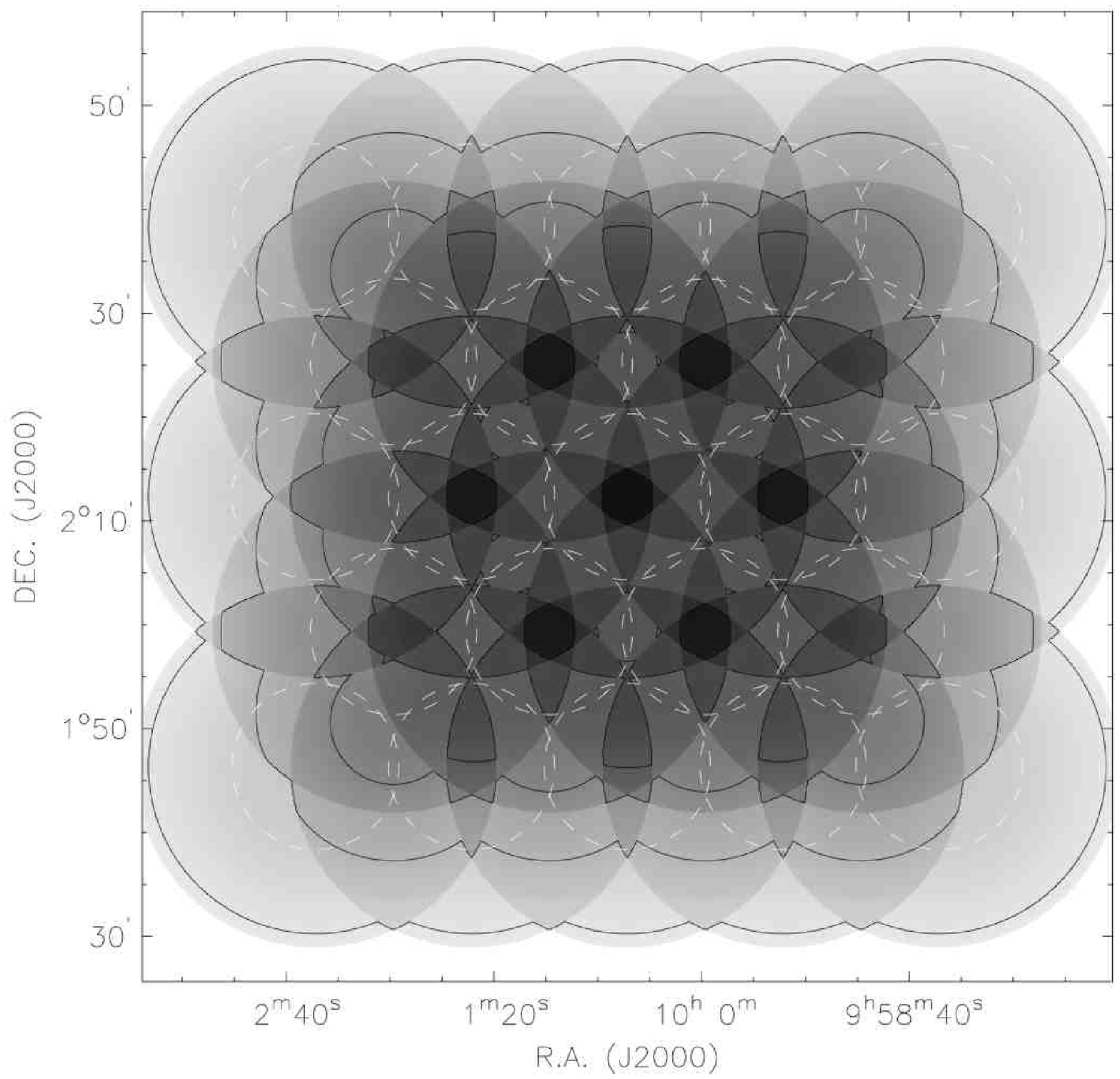} 
\caption{Model map used for the correction of the bandwidth smearing effect. 
The model map is based on the sensitivity drop due to the shape of the primary
beam response. The intensity scale is normalized to the sensitivity at the center
of the field. The contour levels are at 0.1, 0.3, 0.5, 0.7 and 0.9. The area used to
extract sources (r\,$\le$\,8$'$) in each individual pointing for the analysis of Section
\ref{subsec:bwsmeth} is indicated with dashed white circles. \label{fig:modelsens}}
\end{figure}

\clearpage

\begin{figure}
\epsscale{1.}
\plotone{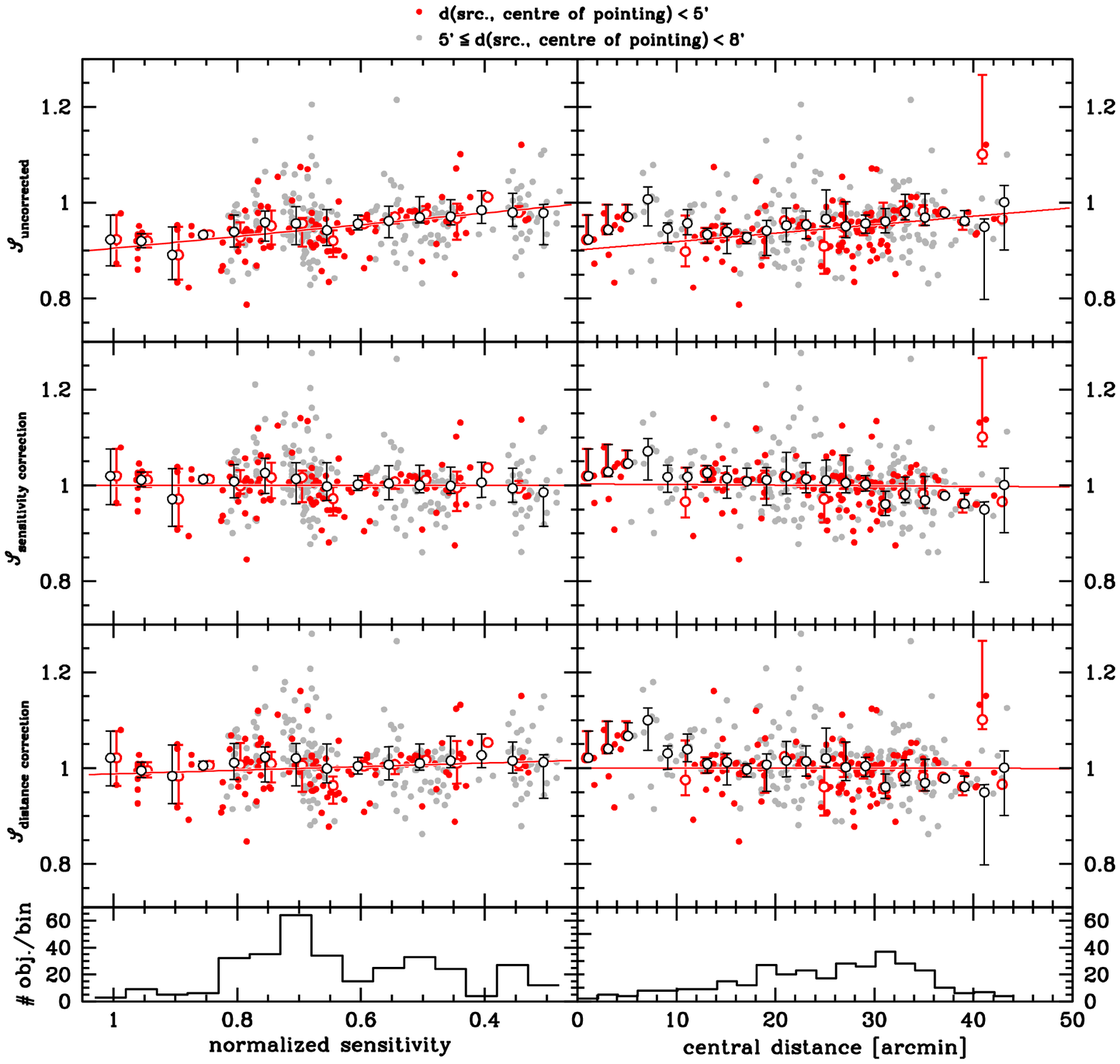} 
\caption{Comparison of two different approaches to correct for the
  bandwidth smearing (BWS) effect. (See text for details)
  \label{fig:bwscorrvar} }
\end{figure}

\clearpage

\begin{figure}
\epsscale{.75}
\plotone{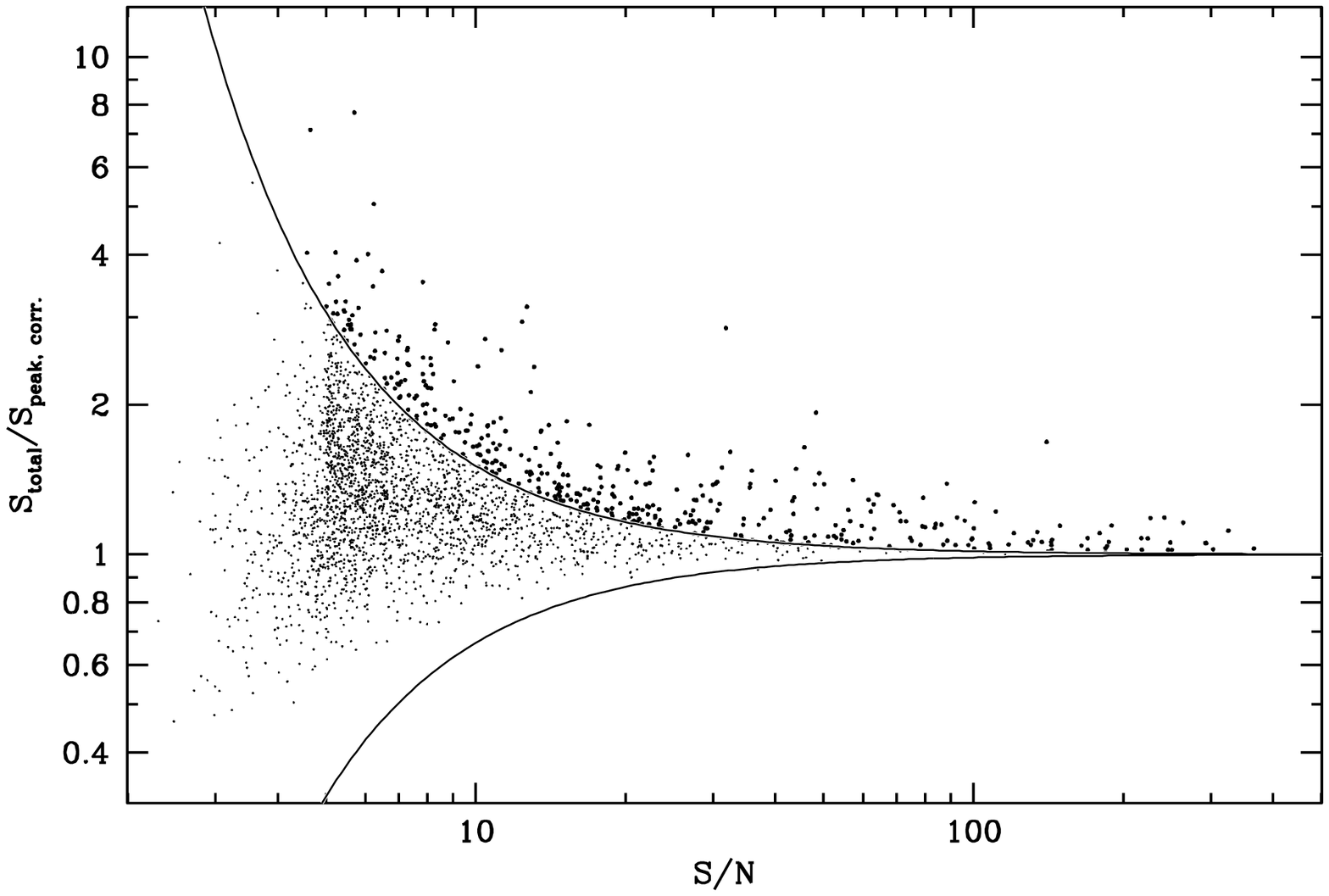}
\caption{Ratio of total $S_{\rm total}$ and 
  peak flux density $S_{\rm peak,\,corr.}$ (BWS corrected)
  as a function of the detection $S/N$, defined as the ratio of the
  (uncorrected) peak flux density and the local (17.5$''$ box) rms. The solid lines
  show the upper and lower envelope of the region defined to contain
  unresolved sources (small dots), based on the simulations presented in the Appendix.
  Sources lying above the upper envelope are considered resolved (large dots).
  \label{fig:resunres}}
\end{figure}

\begin{figure}
\epsscale{.75}
\plotone{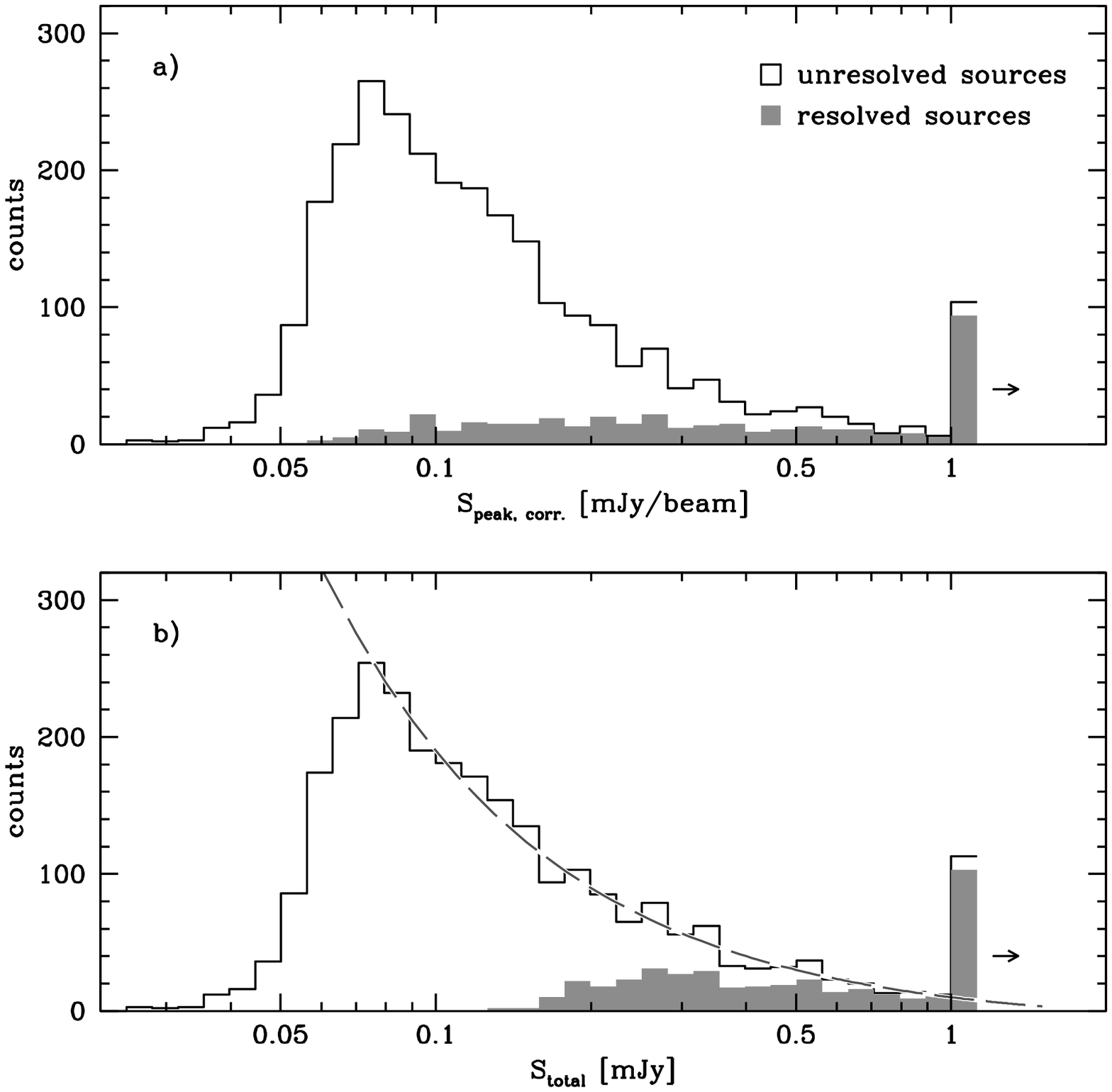} 
\caption{Number distribution of unresolved
  and resolved sources in the VLA-COSMOS Joint catalog as a function
  of corrected peak ({\it top}) and integrated flux density ({\it bottom}).
  Sources with $S_{\rm peak,\,corr.}$ or $S_{\rm total} >$ 1 mJy are
  added to the right-most bin of the histograms. In panel ({\it b}) the
  dashed line (normalized to the counts at 0.1\,mJy) illustrates the
  decline of the integral sources counts $\sim$$S_{\rm total}^{-1}$
  which is expected from the flat part of the Euclidan source counts 
  present at low flux levels \citep[e.g.][]{bon08}.
  \label{fig:flxdistribs}}
\end{figure}

\clearpage

\begin{figure}
\epsscale{.80}
\plotone{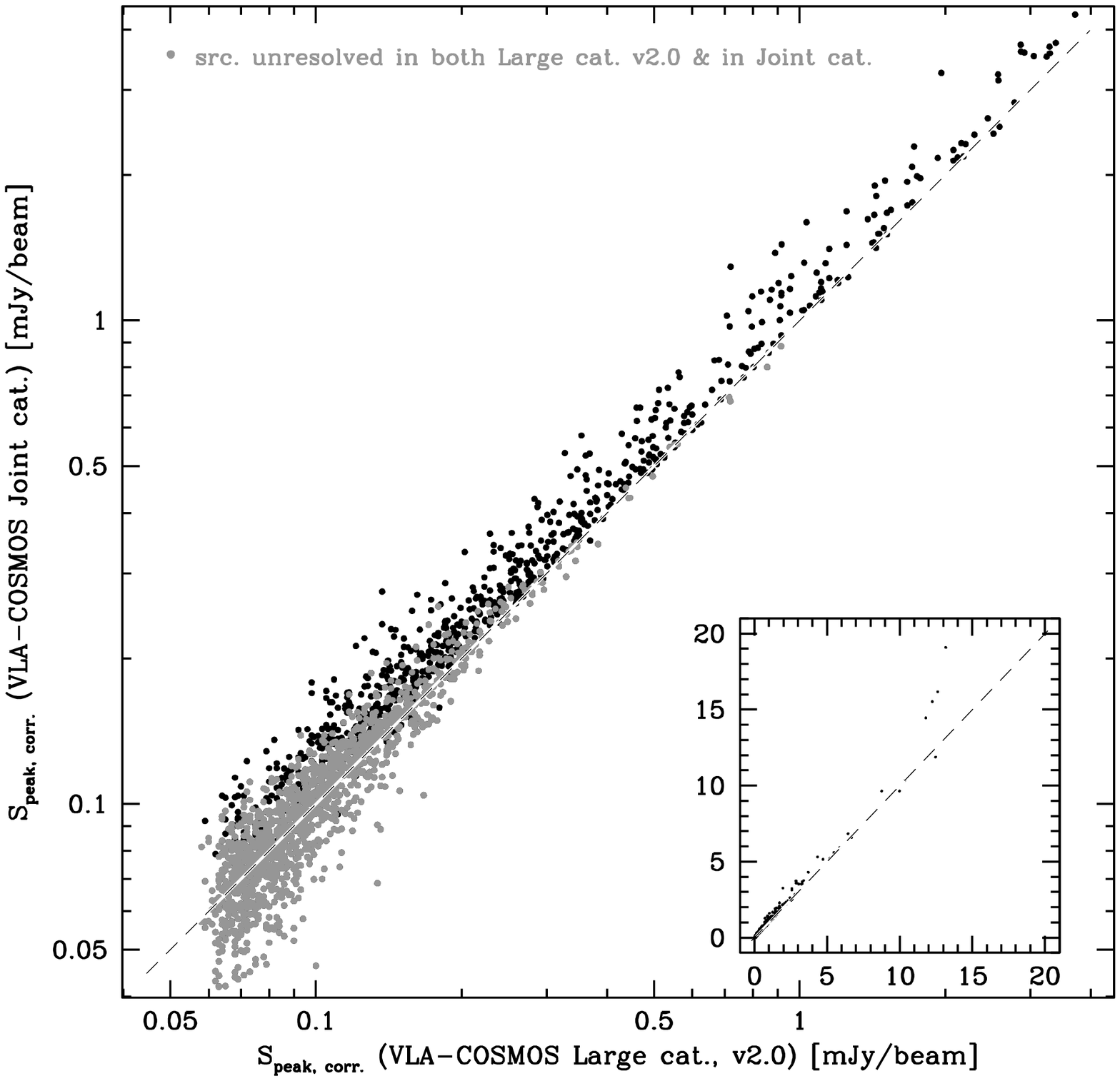} 
\caption{Comparison of the peak flux densities $S_{\rm
    peak}$ (both BWS corrected) in
  the VLA-COSMOS Large (v2.0) and Joint catalog. The inset covers
  the entire range of peak flux densities while the main part of the
  figure is limited to peak flux densities less than 4 mJy/beam. The
  measured values of sources classified as unresolved in both the
  Large (v2.0) as well as the Joint catalog are in good
  agreement (grey dots). Sources classified as
  resolved in the Joint catalog tend to have larger values, as the Deep
  image has a higher sensitivity to extended emission.
  \label{fig:peakcomp}}
\end{figure}

\clearpage

\begin{figure}
\epsscale{.80}
\plotone{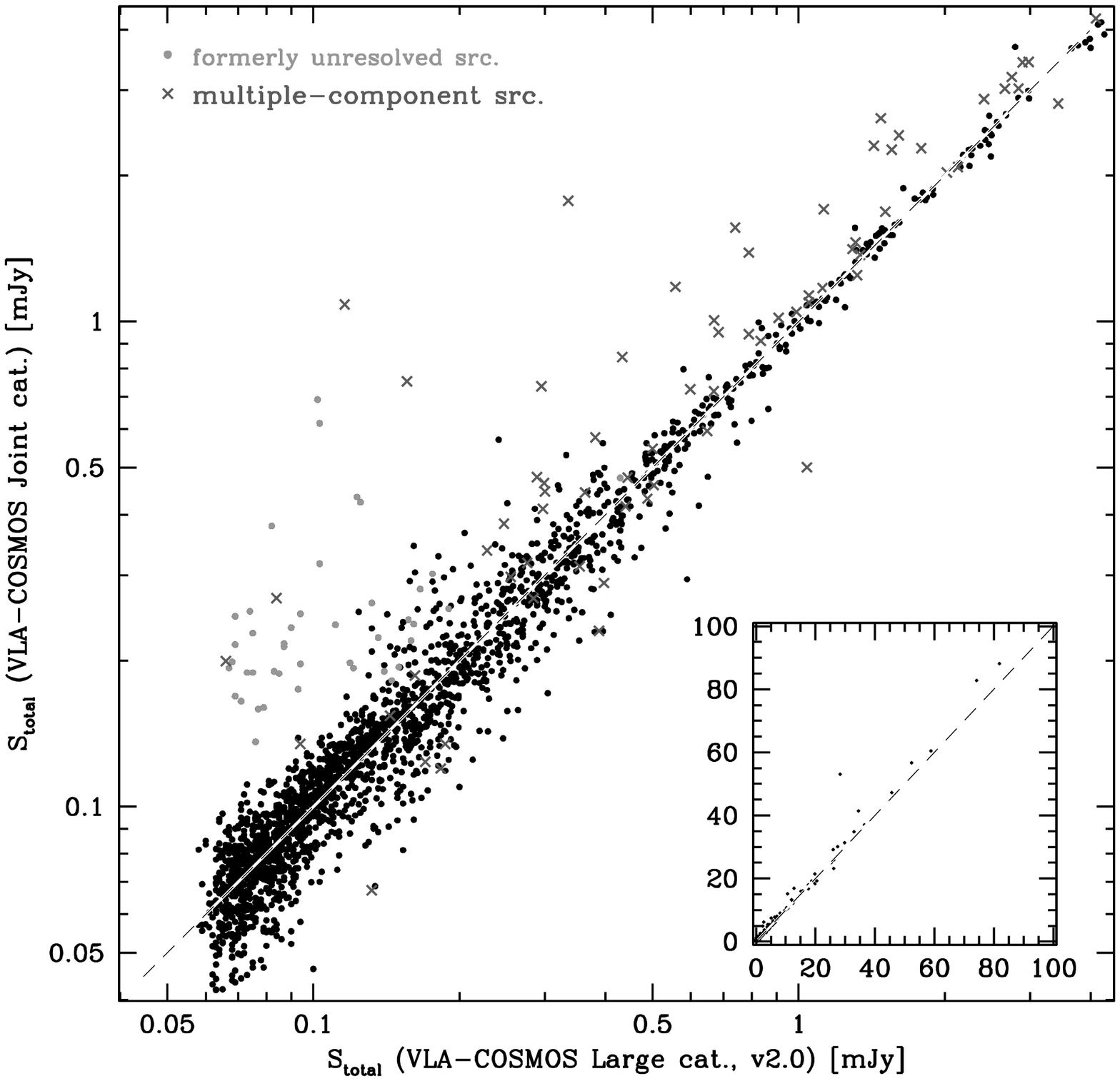} 
\caption{Comparison of the integrated flux $S_{\rm
    total}$ in the VLA-COSMOS Large (v2.0) and Joint catalog (the
  layout of the figure matches that of \ref{fig:peakcomp}). 
   The majority of the multi-component sources (dark grey crosses) 
  have larger measured integrated fluxes in the Joint catalog
  as the Deep mosaic is more sensitive to extended emission. Most
  other sources lie along the diagonal bisector of the plot 
(indicating a good correspondence between the catalogs), except for
  sources that were previously classified as unresolved in the 
   Large catalog (v2.0) and are now classified as 
    resolved in the VLA-COSMOS Joint catalog (light grey points). 
  \label{fig:totcomp}}
\end{figure}

\clearpage

\begin{figure}
\epsscale{.60}
\plotone{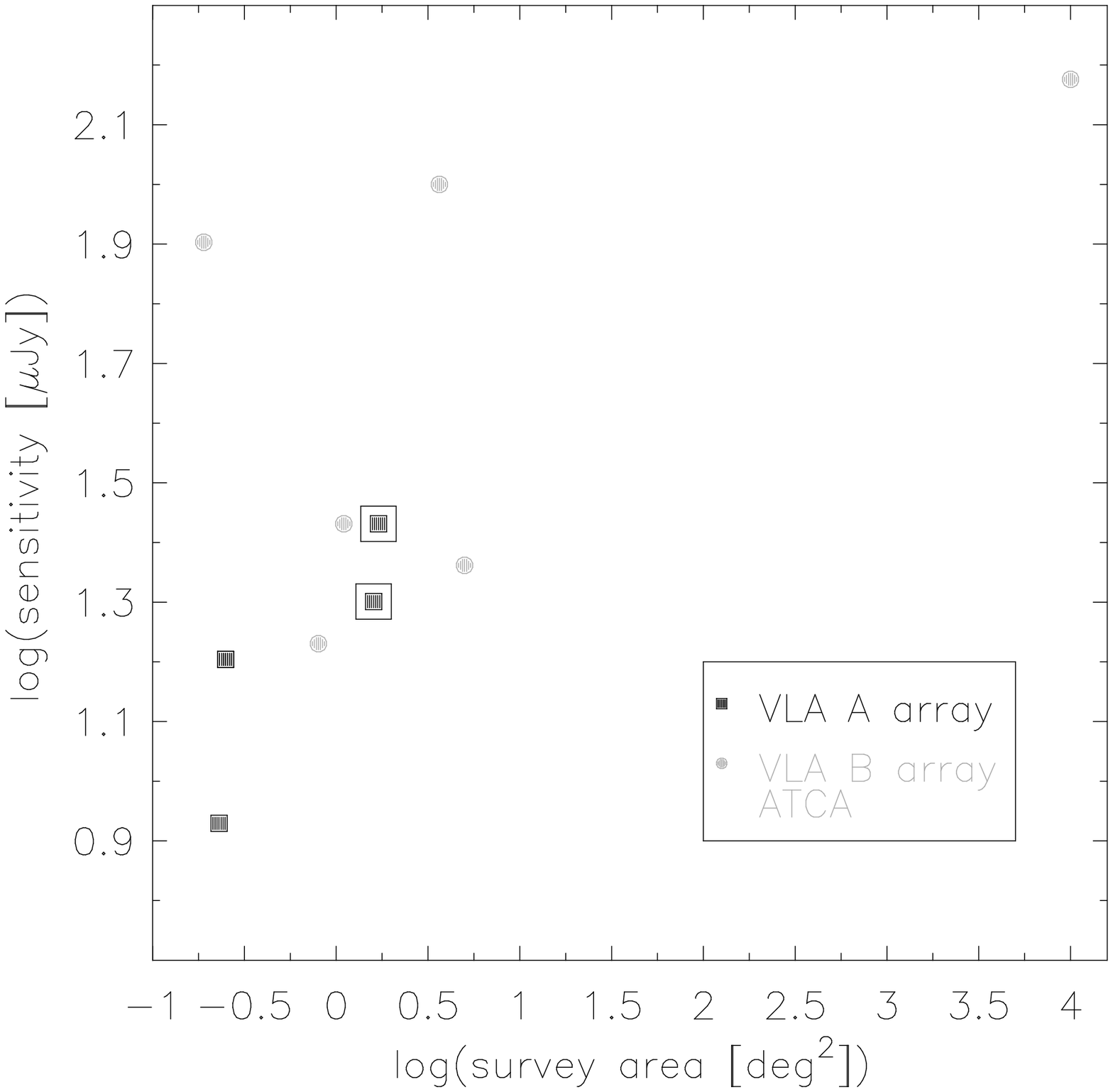} 
\caption{Comparison of representative radio surveys conducted at 20\,cm
  with the VLA and ATCA. Filled squares denote surveys that were
  conducted in the VLA A configuration reaching resolutions of
  $\approx$2$''$, while surveys having resolutions of 5-10$''$ (from the
  VLA B array and ATCA) are shown as filled circles. The VLA-COSMOS
  surveys are marked by open squares (Deep: top, Large: bottom),
  showing that they cover the largest area at their resolution and
  sensitivity.  In order to reasonably compare the different surveys,
  we used the sensitivity that is achieved for at least 80\% of the
  area covered. Thus in the case of surveys that have a non-uniform
  rms pattern the sensitivity value used here can be significantly
  higher than the best value present in the deepest part of the images
  (for detailed numbers see Tab. \ref{tab:survey}).  
  \label{fig:radio_surveys}}
\end{figure}

\clearpage

\begin{figure}
\epsscale{.70}
\plotone{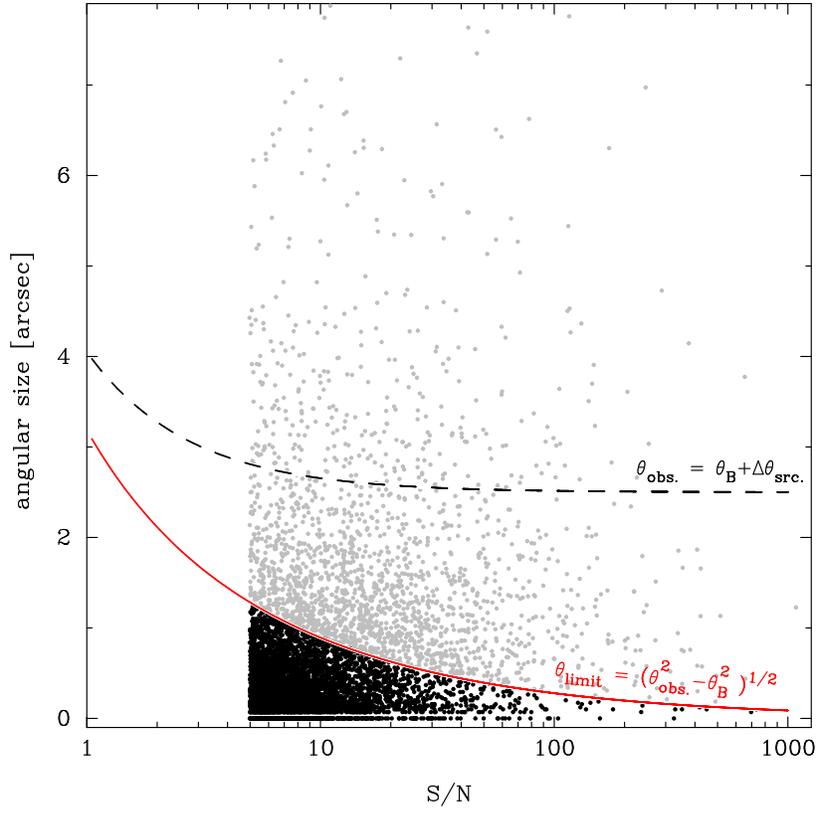}
\caption{Intrinsic source sizes of simulated Gaussian sources vs. 
 their detection
 $S/N$. The dashed line marks the upper expectation value $\theta_{\rm obs.}$
 of the observed size of a point source of varying $S/N$ in the Deep mosaic. 
At
 low $S/N$, point sources may be significantly larger than the beam size, 
$\theta_{\rm B}$.
 The intrinsic (deconvolved) source size, $\theta_{\rm limit}$, inferred 
from the
 dashed line is shown in red. It defines an upper limit to the intrinsic 
source size
 below which a conclusive classification as resolved or unresolved is no longer 
possible.
 Potentially resolvable simulated sources are plotted in grey for easier 
identification
 in Fig. \ref{fig:sim_resunres}.
 \label{fig:Condon_resunres}}
\end{figure}

\clearpage

\begin{figure}
\epsscale{.85}
\plotone{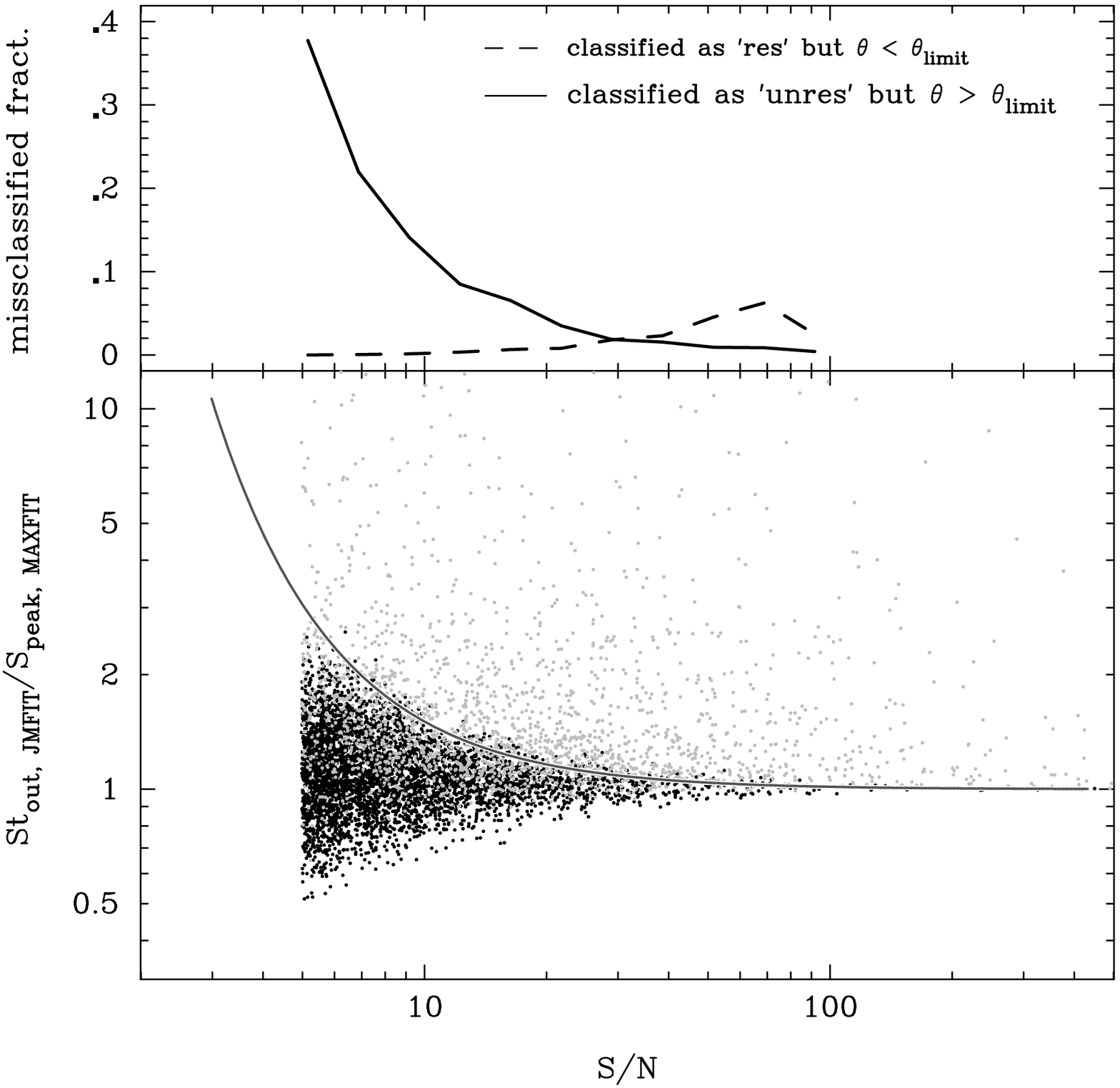}
\caption{Diagnostic diagrams for the separation of simulated Gaussian sources into
 resolved and unresolved objects based on their location in Fig.
 \ref{fig:Condon_resunres} (see text for details). The color of the dots matches that
 used in Fig. \ref{fig:Condon_resunres}.\newline
 In the upper panel we show the fraction of sources in a given bin of $S/N$ that
 (1) have an intrinsic source size $\theta$\,$<$\,$\theta_{\rm limit}$ but which
 nevertheless come to lie above the curve -- given by eq. (\ref{equ2app}; see lower
 panel) -- separating resolved (above curve) from unresolved (below curve) sources,
 or (2) have $\theta$\,$>$\,$\theta_{\rm limit}$ but are found in the area attributed to
 unresolved objects. (The curves are discontinued at $S/N$\,$=$\,100 because small
 number fluctuations dominate beyond this value.)
\label{fig:sim_resunres}}
\end{figure}

\clearpage

\begin{figure}
\epsscale{.8}
\plotone{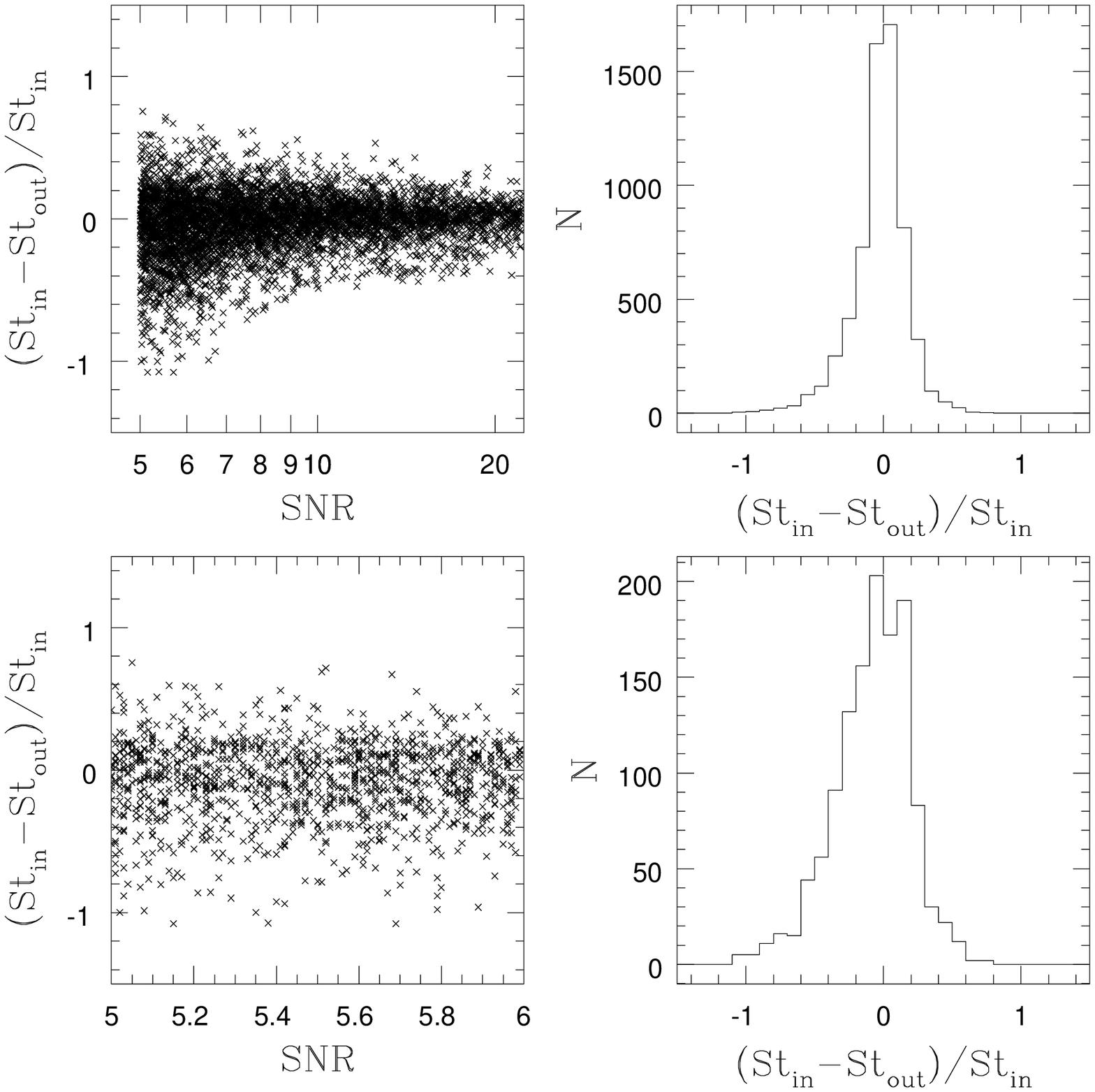}
\caption{Comparison of the recovered total flux density $St_{\rm out}$ of simulated
 Gaussian sources with the input value, $St_{\rm in}$. {\it Top row}: distribution of
 relative errors ({\it right}) and plot of the relative errors as a function of signal-to-noise
 ratio (SNR; {\it left}) for all simulated sources after these have been classified as resolved
 or unresolved using eq. (\ref{equ2app}). (See also Fig. \ref{fig:sim_resunres}.)
 {\it Bottom row}: close-up of the simulated sources in the bin with the smallest
 simulated $S/N$.\newline
 Total flux densities tend to be overestimated by an amount that increases towards
 lower $S/N$. The according average bias is listed in Tab. \ref{tab:comp_inout} for
 resolved sources as well as for the complete sample of resolved and unresolved
 simulated sources.
 \label{fig:comp_inout}}
\end{figure}

\clearpage

\begin{deluxetable}{lrr}
\tablecaption{VLA Pointing Centers for VLA-COSMOS Deep Project\label{tab:pos}}
\tablewidth{0pt}
\tablehead{
\colhead{Pointing \#} & 
\colhead{R.A. (J2000)} & 
\colhead{DEC (J2000)}}
\startdata
F07 & 10:00:58.62 & +02:25:20.42 \\
F08 & 09:59:58.58 & +02:25:20.42 \\
F11 & 10:01:28.64 & +02:12:21.00 \\
F12\tablenotemark{a} & 10:00:28.60 & +02:12:21.00\\
F13 & 09:59:28.56 & +02:12:21.00 \\
F16 & 10:00:58.62 & +01:59:21.58 \\
F17 & 09:59:58.58 & +01:59:21.58 \\
\enddata
\tablenotetext{a}{COSMOS field center} 
\tablecomments{The naming convention of the VLA-COSMOS Large project
  \citep{sch07} has been kept for the pointing centers of the
  VLA-COSMOS Deep project at 1.4\,GHz.}
\end{deluxetable}

\begin{deluxetable}{llrl}
\tabletypesize{\scriptsize}
\tablecaption{Source Numbers in the VLA-COSMOS catalogs\label{tab:vlacos}}
\tablewidth{0pt}
\tablehead{
\colhead{Catalog} & 
\colhead{Type} &
\colhead{\# of objects} & 
\colhead{Remarks}
}
\startdata
Deep Project & components\tablenotemark{a} & 3744 & 4$\sigma$ ($\sim$45\,$\mu$Jy); total\\
Deep Project & sources\tablenotemark{b} & 3441 & 4$\sigma$ ($\sim$45\,$\mu$Jy); total\\ \hline
Large Project (v1.0) & sources & 3643 & 4.5$\sigma$; total\\
& &  1226 & $<5\sigma$\\ \hline
Large Project (v2.0) & sources & 2417 & 5$\sigma$ ($\sim$60\,$\mu$Jy); total\\
& & 1611 & unresolved sources\\
& & 806 & resolved sources \\
& &  78 & multi-component sources \\ \hline
Joint & sources & 2865 & total; combined Large (v2.0) and Deep Project\\
&  & 2159 & detected in both Large (v1.0) and Deep image\\
&  &  392 & from Large Project (v2.0) catalog\\
&  &  314 & from Deep Project source catalog\\
&  & 2329 & unresolved sources \\
&  & 405 & resolved sources \\
&  & 131 & multi-component sources (incl. 43 new ones)\\
\enddata

\tablenotetext{a}{Direct product of running the AIPS task \texttt{SAD} on the Deep image}
\tablenotetext{b}{Cleaned for multi-component and 'twin' sources}
\tablecomments{See text for details. }
\end{deluxetable}

\clearpage

\begin{landscape}
\begin{deluxetable}{lccccccc}
\tabletypesize{\scriptsize}
\tablecaption{1.4\,GHz Joint Source Catalog of the VLA-COSMOS Project (abridged)\label{tab:jointcat}}
\tablewidth{0pt}
\setlength{\tabcolsep}{0.04in}
\tablehead{
\colhead{Name} &
\colhead{former Name} &
\colhead{R.A.} & 
\colhead{Dec.} &
\colhead{R.A.} &
\colhead{Dec.} &
\colhead{$\rm \sigma_{R.A.}$} &
\colhead{$\rm \sigma_{Dec.}$}
\\
& \colhead{(in Large Project catalog, v2)} &
\colhead{[deg], (J2000.0)} &
\colhead{[deg], (J2000.0)} &
\colhead{(J2000.0)} & 
\colhead{(J2000.0)} & 
\colhead{[$\as$]} &
\colhead{[$\as$]}
}
\startdata
COSMOSVLADP\_J095821.65+024628.1 & COSMOSVLA\_J095821.65+024628.1 & 149.5902208 & 2.7744972 & 09 58 21.653 & +02 46 28.19 & 0.25 & 0.25\\
COSMOSVLADP\_J095821.78+024820.6 & COSMOSVLA\_J095821.78+024820.6 & 149.5907833 & 2.8057333 & 09 58 21.788 & +02 48 20.64 & 0.35 & 0.36\\
COSMOSVLADP\_J095821.81+014550.7 & COSMOSVLA\_J095821.82+014550.8 & 149.5908875 & 1.7640944 & 09 58 21.813 & +01 45 50.74 & 0.38 & 0.41\\
COSMOSVLADP\_J095821.82+014724.1 & COSMOSVLA\_J095821.81+014724.2 & 149.5909333 & 1.7900389 & 09 58 21.824 & +01 47 24.14 & 0.26 & 0.26\\
COSMOSVLADP\_J095821.94+020707.7 & COSMOSVLA\_J095821.94+020707.7 & 149.5914333 & 2.1188167 & 09 58 21.944 & +02 07 07.74 & 0.27 & 0.27\\
COSMOSVLADP\_J095822.11+014058.9 & COSMOSVLA\_J095822.10+014058.7 & 149.5921333 & 1.6830278 & 09 58 22.112 & +01 40 58.90 & 0.29 & 0.27\\
COSMOSVLADP\_J095822.18+014524.3 & COSMOSVLA\_J095822.18+014524.3 & 149.5924333 & 1.7567500 & 09 58 22.184 & +01 45 24.30 & 0.29 & 0.30\\
COSMOSVLADP\_J095822.25+013512.3 & COSMOSVLA\_J999999.99+999999.9 & 149.5927083 & 1.5867500 & 09 58 22.250 & +01 35 12.30 & 0.44 & 0.38\\
COSMOSVLADP\_J095822.30+024721.3 & COSMOSVLA\_J095822.30+024721.3 & 149.5929250 & 2.7892583 & 09 58 22.302 & +02 47 21.33 & -99.00 & -99.00\\
COSMOSVLADP\_J095822.57+020239.1 & COSMOSVLA\_J095822.57+020239.1 & 149.5940667 & 2.0441972 & 09 58 22.576 & +02 02 39.11 & 0.27 & 0.29\\
COSMOSVLADP\_J095822.81+023604.3 & COSMOSVLA\_J095822.81+023604.5 & 149.5950583 & 2.6012000 & 09 58 22.814 & +02 36 04.32 & 0.48 & 0.43\\
COSMOSVLADP\_J095822.93+022619.8 & COSMOSVLA\_J095822.93+022619.8 & 149.5955667 & 2.4388333 & 09 58 22.936 & +02 26 19.80 & -99.00 & -99.00\\
COSMOSVLADP\_J095823.25+020859.4 & COSMOSVLA\_J095823.25+020859.4 & 149.5969083 & 2.1498417 & 09 58 23.258 & +02 08 59.43 & -99.00 & -99.00\\
COSMOSVLADP\_J095823.27+021455.9 & COSMOSVLA\_J095823.27+021455.5 & 149.5969917 & 2.2488639 & 09 58 23.278 & +02 14 55.91 & 0.39 & 0.35\\
COSMOSVLADP\_J095823.67+021201.4 & COSMOSVLA\_J095823.68+021201.4 & 149.5986333 & 2.2004000 & 09 58 23.672 & +02 12 01.44 & 0.29 & 0.31\\
COSMOSVLADP\_J095824.02+024916.0 & COSMOSVLA\_J095824.02+024916.0 & 149.6000833 & 2.8211167 & 09 58 24.020 & +02 49 16.02 & -99.00 & -99.00\\
COSMOSVLADP\_J095824.02+025029.5 & COSMOSVLA\_J095824.02+025029.3 & 149.6000875 & 2.8415333 & 09 58 24.021 & +02 50 29.52 & 0.26 & 0.26\\
COSMOSVLADP\_J095824.13+013836.6 & COSMOSVLA\_J095824.14+013836.6 & 149.6005542 & 1.6435250 & 09 58 24.133 & +01 38 36.69 & 0.29 & 0.31\\

\enddata

\tablecomments{VLA-COSMOS Joint Catalog of radio sources at 1.4\,GHz
  representing all reliable radio sources from the Large catalog and
  a search for sources in the Deep mosaic (see text for details). All
  measurements were made on the 2.5$\as\times$2$\as$ Deep image. The
  complete table is available via the link to a machine-readable
  version above and/or at the COSMOS archive at
  IPAC/IRSA\footnote{~\tt{\url{http://www.irsa.ipac.edu/data/COSMOS/tables/vla/}}}.
  A portion is shown here as an example of its form and content.}

\end{deluxetable}
\clearpage
\end{landscape}

\addtocounter{table}{-1}

\begin{deluxetable}{ccccccccccc}
\tabletypesize{\scriptsize}
\tablecaption{{\it cont.}}
\tablewidth{0pt}
\setlength{\tabcolsep}{0.04in}
\tablehead{
\colhead{$\rm S_{peak}$} & 
\colhead{$\rm S_{peak, corr.}\tablenotemark{a}$} &
\colhead{$\rm S_{total}$} &
\colhead{rms} & 
\colhead{$\rm \theta_{M,\,dec}$} &
\colhead{$\rm \theta_{m,\,dec}$} &
\colhead{$\rm PA_{dec}$} &
\colhead{Flags}& & &
\\
\colhead{[mJy/beam]} & 
\colhead{[mJy/beam]} & 
\colhead{[mJy]} &  
\colhead{[mJy/beam]} &
\colhead{[$\as$]} & 
\colhead{[$\as$]} &
\colhead{[$^\circ$]} & 
\colhead{res\tablenotemark{b}}&
\colhead{mult\tablenotemark{c}}&
\colhead{memb\tablenotemark{d}}&
\colhead{det\tablenotemark{e}}
}
\startdata
5.218$\pm$0.029 & 5.218   & 5.218$\pm$0.029 & 0.029 &  0.00 &  0.00 &  0.00 & -1 & 0 & 0 &  0\\
0.201$\pm$0.035 & 0.201   & 0.201$\pm$0.035 & 0.035 &  0.00 &  0.00 &  0.00 &  0 & 0 & -1 & 0\\
0.079$\pm$0.018 & 0.081   & 0.081$\pm$0.018 & 0.018 &  0.00 &  0.00 &  0.00 &  0 & 0 & 0 & -1\\
0.267$\pm$0.014 & 0.274   & 0.328$\pm$0.034 & 0.014 &  1.43 &  0.97 & 60.80 &  1 & 0 & 0 &  0\\
0.179$\pm$0.016 & 0.182   & 0.182$\pm$0.016 & 0.016 &  0.00 &  0.00 &  0.00 &  0 & 0 & 0 &  0\\
0.154$\pm$0.016 & 0.154   & 0.154$\pm$0.016 & 0.016 &  0.00 &  0.00 &  0.00 &  0 & 0 & 0 &  0\\
0.181$\pm$0.017 & 0.185   & 0.341$\pm$0.057 & 0.017 &  3.04 &  1.99 & 94.80 &  1 & 0 & 0 &  0\\
0.135$\pm$0.026 & 0.135   & 0.135$\pm$0.026 & 0.026 &  0.00 &  0.00 & 0.00 &  -2 & 0 & 1 &  1\\
-99.000         & -99.000 & 30.120          & 0.083 & 35.97 &  9.90 &-99.00 &  1 & 1 & 0 &-99\\
0.153$\pm$0.017 & 0.157   & 0.157$\pm$0.017 & 0.017 &  0.00 &  0.00 &  0.00 &  0 & 0 & 0 &  0\\
0.093$\pm$0.019 & 0.096   & 0.096$\pm$0.019 & 0.019 &  0.00 &  0.00 & 0.00 &  0 & 0 & 0 & -1\\
-99.000         & -99.000 & 116.500         & 0.033 & 74.99 & 12.92 &-99.00 &  1 & 1 & 0 &-99\\
-99.000         & -99.000 & 5.025           & 0.026 &  9.90 &  3.75 &-99.00 &  1 & 1 & 0 &-99\\
0.077$\pm$0.016 & 0.079   & 0.079$\pm$0.016 & 0.016 &  0.00 &  0.00 &  0.00 &  0 & 0 & 0 & -1\\
0.162$\pm$0.021 & 0.169   & 0.162$\pm$0.021 & 0.021 &  0.00 &  0.00 &  0.00 & -1 & 0 & 0 &  0\\
-99.000         & -99.000 & 56.620          & 0.067 & 56.98 & 13.93 &-99.00 &  1 & 1 & 0 &-99\\
0.461$\pm$0.029 & 0.461   & 0.461$\pm$0.029 & 0.029 &  0.00 &  0.00 &  0.00 & -1 & 0 & 0 &  0\\
0.141$\pm$0.018 & 0.141   & 0.141$\pm$0.018 & 0.018 &  0.00 &  0.00 &  0.00 &  0 & 0 & 0 &  0\\
\enddata

\tablenotetext{a}{Peak flux value corrected for the bandwidth smearing
  effect.}  
\tablenotetext{b}{Flag for resolved sources (based on Fig.
  \ref{fig:resunres}): (-2) if unresolved only in VLA-COSMOS Deep
  image; (-1) if resolved only in VLA-COSMOS Large image; (0) if
  unresolved in both VLA-COSMOS Large and Deep images; (1) if resolved
  in both VLA-COSMOS Large and Deep images; (2) if resolved only in
  VLA-COSMOS Deep image. Note that the flag value of `-2' is only
  assigned to sources that were not in the VLA-COSMOS Large Project
  catalog. The flag `-1', on the other hand, is used for sources that
  were classified as resolved in the Large Project catalog but are
  listed as unresolved in the Joint catalog.}  
\tablenotetext{c}{Flag
  for distinction between sources consisting of multiple components
  (1) or a single component (0).}  
\tablenotetext{d}{Flag identifying
  whether the source was only detected in the VLA-COSMOS Large image
  (-1), in both the VLA-COSMOS Large and Deep image (0), or only in
  the VLA-COSMOS Deep image (1).}
\tablenotetext{e}{Flag specifying if the source was detected with $S/N_{17.5''}\geq$ 5
only at 1.5$''$ resolution (-1); with $S/N_{17.5''}\geq$ 5 at a resolution of both 1.5$''$
and 2.5$''$ (0); or only at a resolution of 2.5$''$, either in both the large and small scale
rms map (1), or only in the small scale rms map (2).}
\end{deluxetable}


\begin{deluxetable}{lccll}
\tabletypesize{\scriptsize}
\tablecaption{Selected Radio Surveys at 1.4\,GHz\label{tab:survey}}
\tablewidth{0pt}
\tablehead{
\colhead{Field} & 
\colhead{Area\tablenotemark{a}} &  
\colhead{rms\tablenotemark{b}} & 
\colhead{Instrument} &
\colhead{Reference} \\
 &
\colhead{[deg$^2$]} &  
\colhead{[$\mu$Jy/beam]} & 
&}
\startdata
VLA-COSMOS (Deep)  &     1.7 &   27 & VLA-A & this paper \\
VLA-COSMOS (Large) &     1.6 &   20 & VLA-A & Schinnerer et al. 2007\\
ECDF-S             &    0.23 & 8.5  & VLA-A & Miller et al. 2008\\
SSA 13             &    0.25 &   16 & VLA-A & Fomalont et al. 2006\\ \hline
FIRST  &10,000 &  150 & VLA-B & Becker et al. 1995\\
FLS    &     5 &   23 & VLA-B & Condon et al. 2003\\
SXDF               &     1.1 &   27 & VLA-A & Simpson et al. 2006 \\
VVDS   &   0.8 &   17 & VLA-B & Bondi et al. 2003\\
ATHDFS &  0.19 &   80 &  ATCA & Norris et al. 2005, Huynh et al. 2005\\
PDS    &  3.65 &  100 &  ATCA & Hopkins et al. 2003\\
\enddata
\tablenotetext{a}{80\% of the area covered by the survey}
\tablenotetext{b}{Highest rms value occurring for 80\% of the area
  covered} 
\tablecomments{Radio surveys displayed in Fig.
  \ref{fig:radio_surveys}. The values listed were derived from figures
  similar to Fig. \ref{fig:arearms} presented in the corresponding
  references, except for The FIRST and FLS surveys where the full
  areas are used.}
\end{deluxetable}

\clearpage

\begin{deluxetable}{lcc}
\tablecaption{Median of relative error on recovered integrated flux densities for simulated Gaussian sources (see Appendix).\label{tab:comp_inout}}
\tablewidth{0pt}
\tablehead{
$S/N$ range & \colhead{$(St_{\rm in}-St_{\rm out})/St_{\rm in}$ ({\it all} sources)} & \colhead{($St_{\rm in}-St_{\rm out})/St_{\rm in}$ ({\it resolved} sources)}
}
\startdata
5-6 & -0.05$\pm$0.25 & -0.17$\pm$0.42\\
6-7 & -0.02$\pm$0.21 & -0.28$\pm$0.39\\
7-8 & 0.01$\pm$0.17 & -0.16$\pm$0.33\\
8-9 & -0.01$\pm$0.17 & -0.19$\pm$0.29\\
9-10 & -0.01$\pm$0.15 & -0.20$\pm$0.23\\
10-15 & 0.01$\pm$0.12 & -0.10$\pm$0.18\\
15-20 & 0.01$\pm$0.10 & -0.08$\pm$0.15\\
\enddata
\tablecomments{Errors span the range (+/-) containing 2/3 of the measurements.}
\end{deluxetable}

\end{document}